\documentclass[sigconf]{acmart}
\usepackage{booktabs} 
\usepackage{graphicx}  
\usepackage{amssymb}
\usepackage{bm}
\usepackage{amsmath}
\usepackage{verbatim}
\usepackage[ruled,vlined]{algorithm2e}
\usepackage{subfigure}
\usepackage{balance}

\setcopyright{rightsretained}

\begin{document}

\copyrightyear{2018}
\acmYear{2018} 
\setcopyright{iw3c2w3}
\acmConference[WWW 2018]{The 2018 Web Conference}{April 23--27, 2018}{Lyon, France}
\acmBooktitle{WWW 2018: The 2018 Web Conference, April 23--27, 2018, Lyon, France}
\acmPrice{}
\acmDOI{10.1145/3178876.3186146}
\acmISBN{978-1-4503-5639-8/18/04}
\fancyhead{}

\title{Aesthetic-based Clothing Recommendation}
\author{Wenhui Yu$^*$}
\thanks{School of Software, Tsinghua National Laboratory for Information Science and Technology}
\affiliation{%
  \institution{Tsinghua University}
  \city{Beijing}
  \state{China}
}
\email{yuwh16@mails.tsinghua.edu.cn}

\author{Huidi Zhang$^*$}
\thanks{* Both authors contributed equally to this work}
\affiliation{%
  \institution{Tsinghua University}
  \city{Beijing}
  \state{China}
}
\email{zhd16@mails.tsinghua.edu.cn}

\author{Xiangnan He}
\affiliation{%
  \institution{National University of Singapore}
  \state{Singapore, 117417} 
}
\email{xiangnanhe@gmail.com}

\author{Xu Chen}
\affiliation{%
  \institution{Tsinghua University}
  \city{Beijing}
  \state{China}
}
\email{xu-ch14@mails.tsinghua.edu.cn}

\author{Li Xiong}
\affiliation{%
  \institution{Emory University}
  \city{Atlanta}
  \state{USA}
}
\email{lxiong@emory.edu}

\author{Zheng Qin$^\dag$}
\thanks{$\dag$ The corresponding author.}
\affiliation{%
  \institution{Tsinghua University}
  \city{Beijing}
  \state{China}
}
\email{qingzh@mail.tsinghua.edu.cn}


\begin{abstract}

Recently, product images have gained increasing attention in clothing recommendation since the visual appearance of clothing products has a significant impact on consumers' decision. Most existing methods rely on conventional features to represent an image, such as the visual features extracted by convolutional neural networks (CNN features) and the scale-invariant feature transform algorithm (SIFT features), color histograms, and so on. Nevertheless, one important type of features, the \emph{aesthetic features}, is seldom considered. It plays a vital role in clothing recommendation since a users' decision depends largely on whether the clothing is in line with her aesthetics, however the conventional image features cannot portray this directly. To bridge this gap, we propose to introduce the aesthetic information, which is highly relevant with user preference, into clothing recommender systems. To achieve this, we first present the aesthetic features extracted by a pre-trained neural network, which is a brain-inspired deep structure trained for the aesthetic assessment task. Considering that the aesthetic preference varies significantly from user to user and by time, we then propose a new tensor factorization model to incorporate the aesthetic features in a personalized manner. We conduct extensive experiments on real-world datasets, which demonstrate that our approach can capture the aesthetic preference of users and significantly outperform several state-of-the-art recommendation methods.
\vspace{-1mm}
\end{abstract}

%
%



\settopmatter{printacmref=false}
\maketitle

\section{Introduction}
\vspace{-1mm}
\begin{figure}[ht]
\setlength{\abovecaptionskip}{2mm}
\centering
\includegraphics[scale = 0.39]{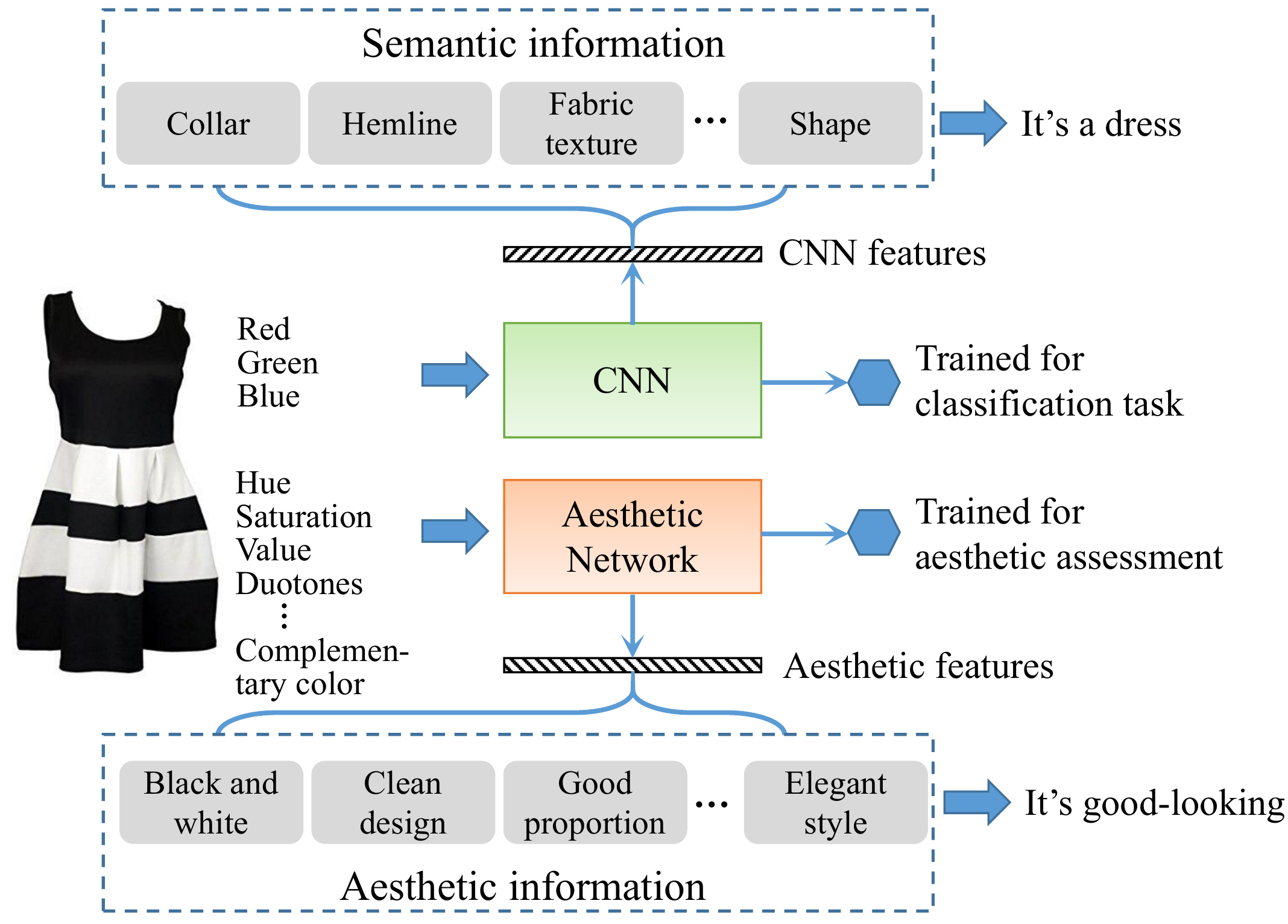}
\caption{Comparison of CNN features and aesthetic features. The CNN is inputted with the RGB components of an image and trained for the classification task, while the aesthetic network is inputted with raw aesthetic features and trained for the aesthetic assessment task.}
\label{fig:Comparison}
\vspace{-2mm}
\end{figure}

When shopping for clothing on the Web, we usually look through product images before making the decision. Product images provide abundant information, including design, color schemes, decorative pattern, texture, and so on; we can even estimate the thickness and quality of a product from its images. As such, product images play a key role in the clothing recommendation task. 

To leverage this information and enhance the performance, existing clothing recommender systems use image data with various image features, like features extracted by convolutional neural networks (CNN features) and the scale-invariant feature transform algorithm (SIFT features), color histograms, etc. For example, \cite{VBPR,Image_based,he_Attentive,Key_Frame} utilized the CNN features extracted by a deep convolutional neural network. Trained for the classification task, CNN features contain semantic information to distinguish items and have been widely used in recommendation tasks. However, one important factor, aesthetics, has yet been considered in previous research. When purchasing clothing products, what consumers concern is not only ``What is the product?'', but also ``Is the product good-looking?''. 

Taking the product shown in Figure \ref{fig:Comparison} as an example. A consumer will notice that the dress is of colors black and white, of simple but elegant design, and has a delightful proportion. She will purchase it only if she is satisfied with all these aesthetic factors. In fact, for some consumers, especially young females, aesthetic factor could be the primary factor, even more important than others like quality, comfort, and prices. As such, we need novel features to capture this indispensable information. Unfortunately, CNN features do not encode the aesthetic information by nature. \cite{matrix} used color histograms to portray consumers' intuitive perception about an image while it is too crude and primitive. To provide quality recommendation for the clothing domain, comprehensive and high-level aesthetic features are greatly desired.


In this paper, we leverage the aesthetic network to extract relevant features. The differences between an aesthetic network and a CNN are demonstrated in Figure \ref{fig:Comparison}.  Recently, \cite{Brain} proposed a \textbf{B}rain-inspired \textbf{D}eep \textbf{N}etwork (\textbf{BDN}), which is a deep structure trained for image aesthetic assessment. The inputs are several raw features that are indicative of aesthetic feelings, like hue, saturation, value, duotones, complementary color, etc. It then extracts high-level aesthetic features from the raw features. In this paper, BDN is utilized to extract the holistic features to represent the aesthetic elements of a clothing product (taking Figure \ref{fig:Comparison} as an example, the aesthetic elements can be color, structure, proportion, style, etc.). 

It is obvious that the aesthetic preference shows a significant diversity among different people. For instance, children prefer colorful and lovely products while adults prefer those can make them look mature and elegant; women may prefer exquisite decorations while men like concise designs. Moreover, the aesthetic tastes of consumers also change with time, either in short term, or in long term. For example, the aesthetic tastes vary in different seasons periodically---in spring or summer, people may prefer clothes with light color and fine texture, while in autumn or winter, people tend to buy clothes with dark color, rough texture, and loose style. In the long term, the fashion trend changes all the time and the popular color and design may be different by year. 

To capture the diversity of the aesthetic preference among consumers and over time, we exploit tensor factorization as a basic model. There are several ways to decompose a tensor \cite{Tensor_application,Pairwise,Tensor_for_Signal_Processing}, however, there are certain drawbacks in existing models. To address the clothing recommendation task better, we first propose a \textbf{D}ynamic \textbf{C}ollaborative \textbf{F}iltering (\textbf{DCF}) model trained with coupled matrices to mitigate the \emph{sparsity} problem \cite{All_at_once}. We then combine it with the additional image features (concatenated aesthetic and CNN features) and term the method as \textbf{D}ynamic \textbf{C}ollaborative \textbf{F}iltering model with \textbf{A}esthetic Features (called \textbf{DCFA}). We optimize the models with bayesian personalized ranking (BPR) optimization criterion \cite{BPR} and evaluated their performance on an \emph{Amazon clothing} dataset. Extensive experiments show that we improve the performance significantly by incorporating aesthetic features. 

To summarize, our main contributions are as follows:

\begin{itemize}
\item{We leverage novel aesthetic features in recommendation to capture consumers' aesthetic preference. Moreover, we compare the effect with several conventional features to demonstrate the necessity of the aesthetic features.}

\item{We propose a novel DCF model to portray the purchase events in three dimensions: users, items, and time. We then incorporate aesthetic features into DCF and train it with coupled matrices to alleviate the sparsity problem.}

\item{We conduct comprehensive experiments on real-world datasets to demonstrate the effectiveness of our DCFA method.}
\end{itemize}

\section{Related Work}

This paper develops aesthetic-aware clothing recommender systems. Specifically, we incorporate the features extracted from the product images by an aesthetic network into a tensor factorization model. As such, we review related work on aesthetic networks, image-based recommendation, and tensor factorization. 

\subsection{Aesthetic Networks}
The aesthetic networks are proposed for image aesthetic assessment. After \cite{Studying_aesthetic} first proposed the aesthetic assessment problem, many research efforts exploited various handcrafted features to extract the aesthetic information of images \cite{Studying_aesthetic,Design_of_High-Level,Content-based,Assessing_the_aesthetic_quality}. To portray the subjective and complex aesthetic perception, \cite{Learning,RAPID,Brain,O6,O8} exploited deep networks to emulate the underlying complex neural mechanisms of human perception, and displayed the ability to describe image content from the primitive level (low-level) features to the abstract level (high-level) features.

\subsection{Image-based Recommendations}
Recommendation has been widely studied due to its extensive use, and many effective methods have been proposed \cite{BPR,HLBPR,O2,O3,R1,R2,R3,R4,NCF,H1,he_Generic,he_item}. The power of recommender systems lies on their ability to model the complex preference that consumers exhibit toward items based on their past interactions and behavior. To extend their expressive power, various works exploited image data \cite{Image_based,VBPR,Key_Frame,matrix,O4,O5,he_Attentive,he_context,he_embedding}. For example, \cite{O4} infused product images and item descriptions together to make dynamic predictions, \cite{Key_Frame,he_context} leveraged textual and visual information to recommend tweets and personalized key frames respectively. Image data can also mitigate the sparsity problem and cold start problem. \cite{Image_based,VBPR,O5,he_Attentive} used CNN features of product images while \cite{matrix} recommended movies with color histograms of posters and frames. \cite{clothes1,clothes2,clothes3} recommended clothes by considering the clothing fashion style.

\subsection{Tensor Factorization}
Time is an important contextual information in recommender systems since the sales of commodities show a distinct time-related succession. In context-aware recommender systems, tensor factorization has been extensively used. For example, \cite{Tensor_application,Tensor_for_Signal_Processing} introduced two main forms of tensor decomposition, the \textbf{C}ANDECOMP/\textbf{P}ARAFAC (\textbf{CP}) and Tucker decomposition. \cite{O10} first utilized tensor factorization for context-aware collaborative filtering. \cite{Pairwise,ChenXu} proposed a \textbf{P}airwise \textbf{I}nteraction \textbf{T}ensor \textbf{F}actorization (\textbf{PITF}) model to decompose the tensor with a linear complexity. Nevertheless, tensor-based methods suffer from several drawbacks like poor convergence in sparse data \cite{poor_convergence} and not scalable to large-scale datasets \cite{Scalable}. To address these limitations, \cite{All_at_once,who,O11} formulated recommendation models with the \textbf{C}oupled \textbf{M}atrix and \textbf{T}ensor \textbf{F}actorization (\textbf{CMTF}) framework.

\begin{figure*}[ht]
\setlength{\abovecaptionskip}{2mm}
\centering
\includegraphics[scale = 0.55]{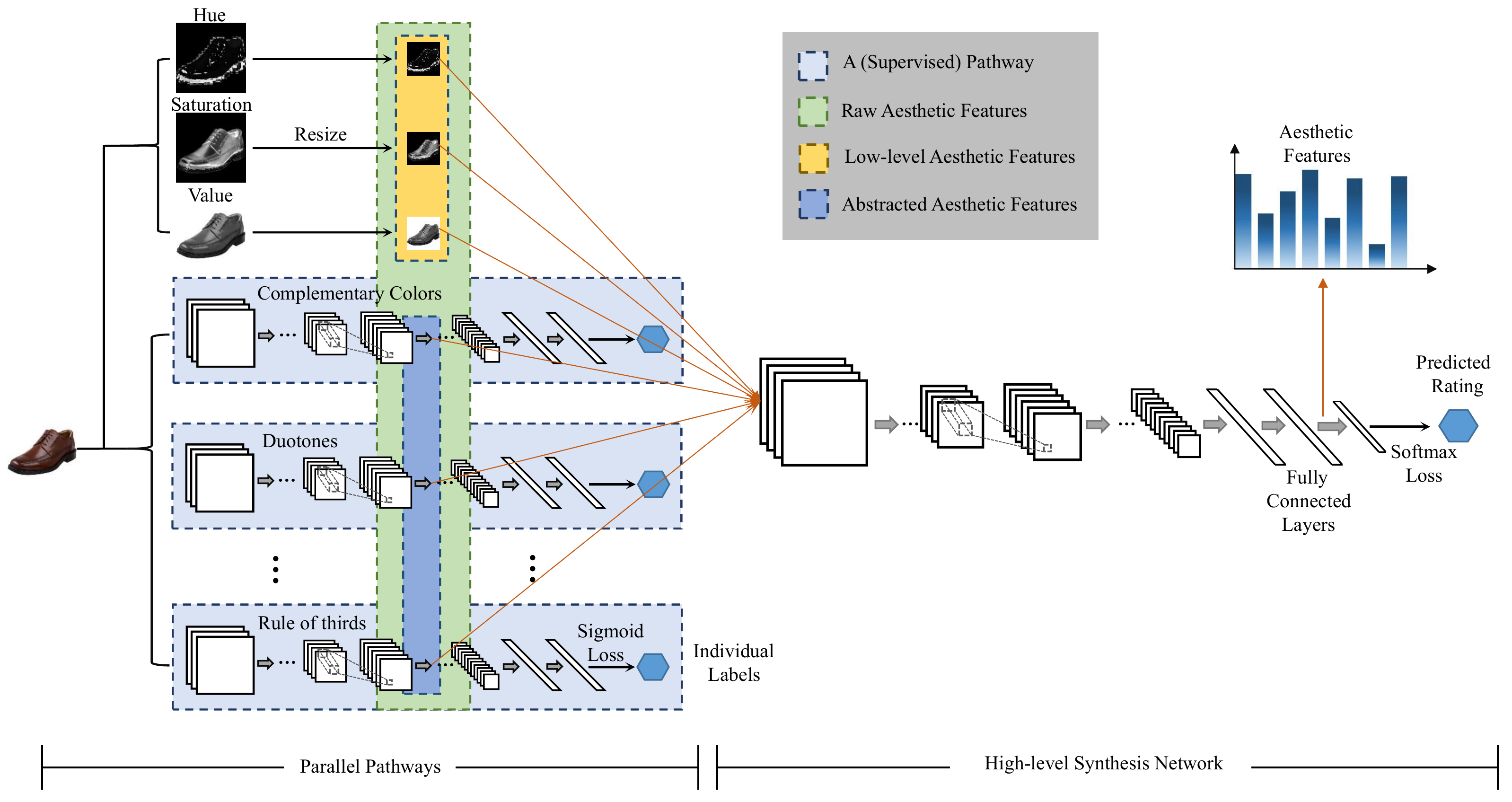}
\caption{Brain-inspired Deep Network (BDN) architecture.}
\label{fig:aesthetic}
\vspace{-2mm}
\end{figure*}

\section{Preliminaries}

This section introduces some preliminaries about the aesthetic neural network, which is used to extract the aesthetic features of clothing images. In \cite{Brain}, the authors introduced the Brain-inspired Deep Networks (BDN, shown in Figure \ref{fig:aesthetic}), a deep CNN structure consists of several parallel pathways (sub-networks) and a high-level synthesis network. It is trained on the \emph{Aesthetic Visual Analysis (AVA)} dataset, which contains 250,000 images with aesthetic ratings and tagged with 14 photographic styles (e.g., complementary colors, duotones, rule of thirds, etc.). The pathways take the form of convolutional networks to exact the abstracted aesthetic features by \emph{pre-trained} with the individual labels of each tag. For example, when training the pathway for complementary colors, the individual label is 1 if the sample is tagged with ``complementary colors'' and is 0 if not. We input the raw features, which include low-level features (hue, saturation, value) and abstracted features (feature maps of the pathways), into the high-level synthesis network and \emph{jointly tune} it with the pathways for aesthetic rating prediction. Considering that the \emph{AVA} is a photography dataset and the styles are for photography, so not all the raw features extracted by the pathways are desired in our recommendation task. Thus we only reserve the pathways that are relevant to the clothing aesthetic. Finally, we use the output of the second fully-connected layer of the synthesis network as our aesthetic features.

We then analyze several extensively used features and demonstrate the superiority of our aesthetic features.

\textbf{CNN Features:} These are the most extensively used features due to their extraordinary representation ability. Typically the output of certain fully-connected layer of a deep CNN structure is used. For example, a common choice is the Caffe reference model with 5 convolutional layers followed by 3 fully-connected layers (pre-trained on the ImageNet dataset); the features are the output of FC7, namely, the second fully-connected layer, which is a feature vector of length 4096. 

CNN features mainly contain semantic information, which contributes little to evaluate the aesthetics of an image. Recall the example in Figure \ref{fig:Comparison}, it can encode ``There is a skirt in the image.'' but cannot express ``The clothing is beautiful and fits the consumer's taste.''. Devised for aesthetic assessment, BDN can capture the high-level aesthetic information. As such, our aesthetic features can do better in beauty estimating and complement CNN features in clothing recommendation.

\textbf{Color Histograms:} \cite{matrix} exploited color histograms to represent human's feeling about the posters and frames for movie recommendation. Though can get the aesthetic information roughly, the low-level handcrafted features are crude, unilateral, and empirical. BDN can get abundant visual features by the pathways. Also, it is data-driven, since the rules to extract features are learned from the data. Compared with the intuitive color histograms, our aesthetic features are more objective and comprehensive. Recall the example in Figure \ref{fig:Comparison} again, color histograms can tell us no more than ``The clothes in the image is white and black''.

\section{Clothing Recommendation with Aesthetic Features}
In this section, we first introduce the basic tensor factorization model (DCF). We next construct a hybrid model that integrates image features into the basic model (DCFA).

\subsection{Basic Model}

Considering the impact of time on aesthetic preference, we propose a context-aware model as the basic model to account for the temporal factor. We use a $P \times Q \times R$ tensor ${\bm{{\rm A}}}$ to indicate the purchase events among the user, clothes, and time dimensions (where $P$, $Q$, $R$ are the number of users, clothes, and time intervals, respectively). If user $p$ purchased item $q$ in time interval $r$, ${\bm{{\rm A}}}_{pqr}=1$, otherwise ${\bm{{\rm A}}}_{pqr}=0$. Tensor factorization has been widely used to predict the missing entries (i.e., zero elements) in ${\bm{{\rm A}}}$, which can be used for recommendation. There are several approaches and we introduce the most common ones:

\subsubsection{Existing Methods and Their Limitations} In this subsection, we summarize the motivation of proposing our novel tensor factorization
model.

\textbf{Tucker Decomposition:} This method \cite{Tensor_application} decomposes the tensor ${\bm{{\rm A}}}$ into a tensor core and three matrices, 
$$\hat{\bm{{\rm A}}}_{pqr}=\sum_{i=1}^{K_1} \sum_{j=1}^{K_2} \sum_{k=1}^{K_3} {\bm{{\rm a}}}_{ijk} {\bm{{\rm U}}}_{ip} {\bm{{\rm V}}}_{jq} {\bm{{\rm T}}}_{kr},$$
where ${\bm{{\rm a}}} \in \mathbb{R}^{K_1 \times K_2 \times K_3}$ is the tensor core, ${\bm{{\rm U}}} \in \mathbb{R}^{K_1 \times P}$, ${\bm{{\rm V}}} \in \mathbb{R}^{K_2 \times Q}$, and ${\bm{{\rm T}}} \in \mathbb{R}^{K_3 \times R}$. Tucker decomposition has very strong representation ability, but it is very time consuming, and hard to converge.

\textbf{CP Decomposition:} The tensor ${\bm{{\rm A}}}$ is decomposed into three matrices in CP decomposition,
$$\hat{\bm{{\rm A}}}_{pqr}=\sum_{k=1}^{K} {\bm{{\rm U}}}_{kp} {\bm{{\rm V}}}_{kq} {\bm{{\rm T}}}_{kr},$$
where ${\bm{{\rm U}}} \in \mathbb{R}^{K \times P}$, ${\bm{{\rm V}}} \in \mathbb{R}^{K \times Q}$, and ${\bm{{\rm T}}} \in \mathbb{R}^{K \times R}$. This model has been widely used due to its linear time complexity, especially in Coupled Matrix and Tensor Factorization (CMTF) structure model \cite{All_at_once,who,Scalable}. However, all dimensions (users, clothes, time) are related by the same latent features. Intuitively, we want the latent features relating users and clothes to contain the information about users' preference, like aesthetics, prices, quality, brands, etc., and the latent features relating clothes and time to contain the information about the seasonal characteristics and fashion elements of clothes like colors, thickness, design, etc. 

\textbf{PITF Decomposition:} The Pairwise Interaction Tensor Factorization (PITF) model \cite{Pairwise} decomposes ${\bm{{\rm A}}}$ into three pair of matrices,
$$\hat{\bm{{\rm A}}}_{pqr}=\sum_{k=1}^{K} {\bm{{\rm U}}}_{kp}^{\bm{{\rm V}}} {\bm{{\rm V}}}_{kq}^{\bm{{\rm U}}} +\sum_{k=1}^{K} {\bm{{\rm U}}}_{kp}^{\bm{{\rm T}}} {\bm{{\rm T}}}_{kr}^{\bm{{\rm U}}} + \sum_{k=1}^{K} {\bm{{\rm V}}}_{kq}^{\bm{{\rm T}}} {\bm{{\rm T}}}_{kr}^{\bm{{\rm V}}},$$
where ${\bm{{\rm U}}}^{\bm{{\rm V}}}, {\bm{{\rm U}}}^{\bm{{\rm T}}} \in \mathbb{R}^{K \times P}$; ${\bm{{\rm V}}}^{\bm{{\rm U}}}, {\bm{{\rm V}}}^{\bm{{\rm T}}} \in \mathbb{R}^{K \times Q}$; ${\bm{{\rm T}}}^{\bm{{\rm U}}}, {\bm{{\rm T}}}^{\bm{{\rm V}}} \in \mathbb{R}^{K \times R}$. PIFT has a linear complexity and strong representation ability. Yet, it is not in line with practical applications due to the additive combination of each pair of matrices. For example, in PIFT, for certain clothes $q$ liked by the user $p$ but not fitting the current time $r$, $q$ gets a high score for $p$ and a low score for $r$. Intuitively it should not be recommended to the user since we want to recommend the right item in the right time. However, the total score can be high enough if $p$ likes $q$ so much that $q$'s score for $p$ is very high. In this case, $q$ will be returned even it does not fit the time. In addition, PITF model is inappropriate to be trained with coupled matrices.  

\subsubsection{Dynamic Collaborative Filtering (DCF) Model}

To address the limitations of the aforementioned models, we propose a new tensor factorization method. When a user makes a purchase decision on a clothing product, there are two primary factors: if the product fits the user's preference and if it fits the time. A clothing product fits a user's preference if the appearance is appealing, the style fits the user's tastes, the quality is good, and the price is acceptable. And a clothing product fits the time if it is in-season and fashionable. For user $p$, clothing $q$, and time interval $r$, we use the scores $S_1$ and $S_2$ to indicate how the user likes the clothing and how the clothing fits the time respectively. $S_1 = 1$ when the user likes the clothing and $S_1 = 0$ otherwise. Similarly, $S_2 = 1$ if the clothing fits the time and $S_2 = 0$ otherwise. The consumer will buy the clothing only if $S_1 = 1$ and $S_2 = 1$, so, $\hat{{\bm{{\rm A}}}}_{pqr} = S_1 \& S_2$. To make the formula differentiable, we can approximately formulate it as $\hat{{\bm{{\rm A}}}}_{pqr} = S_1 \cdot S_2$. We present $S_1$ and $S_2$ in the form of matrix factorization: $$S_1 = \sum_{i = 1}^{K_1} {\bm{{\rm U}}}_{ip} {\bm{{\rm V}}}_{iq}$$ $$S_2 = \sum_{j = 1}^{K_2} {\bm{{\rm T}}}_{jr} {\bm{{\rm W}}}_{jq},$$ where ${\bm{{\rm U}}} \in \mathbb{R}^{K_1 \times P}$, ${\bm{{\rm V}}} \in \mathbb{R}^{K_1 \times Q}$, ${\bm{{\rm T}}} \in \mathbb{R}^{K_2 \times R}$, and ${\bm{{\rm W}}} \in \mathbb{R}^{K_2 \times Q}$. The prediction is then given by:
\begin{eqnarray}
\label{equ:newpair}
\hat{{\bm{{\rm A}}}}_{pqr} = \left({\bm{{\rm U}}}_{*p}^\mathsf{T} {\bm{{\rm V}}}_{*q}\right) \left({\bm{{\rm T}}}_{*r}^\mathsf{T} {\bm{{\rm W}}}_{*q}\right).
\end{eqnarray}
We can see that in Equation (\ref{equ:newpair}), the latent features relating users and clothes are independent with those relating clothes and time. Though $K_1$-dimensional vector ${\bm{{\rm V}}}_{*q}$ and $K_2$-dimensional vector ${\bm{{\rm W}}}_{*q}$ are all latent features of clothing $q$, ${\bm{{\rm V}}}_{*q}$ captures the information about users' preference intuitively whereas ${\bm{{\rm W}}}_{*q}$ captures the temporal information of the clothing. Compared with CP decomposition, our model is more expressive in capturing the underlying latent patterns in purchases. Compared with PITF, combining $S_1$ and $S_2$ with \& (approximated by multiplication) is helpful to recommend right clothing in right time. Moreover, our model is efficient and easy to train compared with the Tucker decomposition.

\subsubsection{Coupled Matrix and Tensor Factorization}
Though widely used to portray the context information in recommendation, tensor factorization suffers from poor convergence due to the sparsity of the tensor. To relieve this problem, \cite{All_at_once} proposed a CMTF model, which decomposes the tensor with coupled matrices. In this subsection, we couple our tensor factorization model with restrained matrices during training.

\textbf{User $\times$ Clothing Matrix:} We use matrix ${\bm{{\rm B}}} \in \mathbb{R}^{P \times Q}$ to indicate the purchase activities between users and clothes. ${\bm{{\rm B}}}_{pq} = 1$ if the user $p$ purchased clothing $q$ and ${\bm{{\rm B}}}_{pq} = 0$ if not.

\textbf{Time $\times$ Clothing Matrix:} We use matrix ${\bm{{\rm C}}} \in \mathbb{R}^{R \times Q}$ to record when the clothing was purchased. Since the characteristics of clothing change steadily with time, we do a coarse-grained discretization on time to avoid the tensor from being extremely sparse. Time is divided into $R$ intervals in total. ${\bm{{\rm C}}}_{rq} = 1$ if the clothing $q$ is purchased in time interval $r$ and ${\bm{{\rm C}}}_{rq} = 0$ if not.

\textbf{Objective Function Formulation:} In existing works \cite{All_at_once,who,O10,O11}, CMTF models are optimized by minimizing the sum of the squared error of each simulation ($\rm MSE\_O\scriptstyle PT$). It is represented as:

\begin{eqnarray}
\label{equ:objective_function}
\left.\begin{aligned}
\rm MSE\_O\scriptstyle PT = &\frac{1}{2} {\left\Arrowvert \bm{{\rm A}} \!-\! \hat{\bm{{\rm A}}} \right\Arrowvert}_{\rm F}^2
\!+\! \frac{\lambda_1}{2} {\left\Arrowvert \bm{{\rm B}} \!-\! \hat{\bm{{\rm B}}} \right\Arrowvert}_{\rm F}^2
\!+\! \frac{\lambda_2}{2} {\left\Arrowvert \bm{{\rm C}} \!-\! \hat{\bm{{\rm C}}} \right\Arrowvert}_{\rm F}^2\\
&+\! \frac{\lambda_3}{2} {\left\Arrowvert \bm{{\rm U}} \right\Arrowvert}_{\rm F}^2
\!+\! \frac{\lambda_4}{2} {\left\Arrowvert \bm{{\rm V}} \right\Arrowvert}_{\rm F}^2
\!+\! \frac{\lambda_5}{2} {\left\Arrowvert \bm{{\rm T}} \right\Arrowvert}_{\rm F}^2
\!+\! \frac{\lambda_6}{2} {\left\Arrowvert \bm{{\rm W}} \right\Arrowvert}_{\rm F}^2,
\end{aligned}
\right.
\qquad
\end{eqnarray}
where $\hat{\bm{{\rm A}}}$ is defined in Equation (\ref{equ:newpair}), $\hat{\bm{{\rm B}}} = \bm{{\rm U}}^\mathsf{T} \bm{{\rm V}}$, $\hat{\bm{{\rm C}}} = \bm{{\rm T}}^\mathsf{T} \bm{{\rm W}}$, and ${\left\Arrowvert \;\; \right\Arrowvert}_{\rm F}$ is the Frobenius norm of the matrix. The last four terms of Equation (\ref{equ:objective_function}) are the regularization terms to prevent overfitting. Although the pointwise squared loss has been widely used in recommendation, it is not directly optimized for ranking. To get better top-$n$  performance, we next introduce our hybrid model with BPR \cite{BPR} optimization criterion.

\subsection{Hybrid Model}

\subsubsection{Problem Formulation}

Combined with image features, we formulate the predictive model as:
\begin{eqnarray}
\label{equ:pair_image}
\hat{{\bm{{\rm A}}}}_{pqr} = \left({\bm{{\rm U}}}_{*p}^\mathsf{T} {\bm{{\rm V}}}_{*q} + {\bm{{\rm M}}}_{*p}^\mathsf{T} {\bm{{\rm F}}}_{*q}\right) \left({\bm{{\rm T}}}_{*r}^\mathsf{T} {\bm{{\rm W}}}_{*q} + {\bm{{\rm N}}}_{*r}^\mathsf{T} {\bm{{\rm F}}}_{*q}\right),
\end{eqnarray}
where ${\bm{{\rm F}}} \in \mathbb{R}^{K \times Q}$ is the feature matrix, ${\bm{{\rm F}}}_{*q}$ is the image features of clothing $q$, which is the concatenation of CNN features (${\bm{{\rm f}}}_{CNN}$) and aesthetic features (${\bm{{\rm f}}}_{AES}$), ${\bm{{\rm F}}}_{*q} = \left[ \begin{matrix} {\bm{{\rm f}}}_{CNN}\\{\bm{{\rm f}}}_{AES} \end{matrix} \right]$ and $K = 8192$. ${\bm{{\rm M}}} \in \mathbb{R}^{K \times P}$ and ${\bm{{\rm N}}} \in \mathbb{R}^{K \times R}$ are aesthetic preference matrices. ${\bm{{\rm M}}}_{*p}$ encodes the preference of user $p$ and ${\bm{{\rm N}}}_{*r}$ encodes the preference in time interval $r$. In our model, both the latent features and image features contribute to the final prediction. Though the latent features can uncover any relevant attribute theoretically, they usually cannot in real-world applications on account of the sparsity of the data and lack of information. So the assistance of image information can highly enhance the model. Also, recommender systems often suffer from the \emph{cold start} problem. It is hard to extract information from users and clothes without consumption records. In this case, content and context information can alleviate this problem. For example, for certain ``cold'' clothing $q$, we can decide whether to recommend it to certain consumer $p$ in current time $r$ according to if $q$ looks satisfying to the consumer (determined by ${\bm{{\rm M}}}_{*p}$) and to the time (determined by ${\bm{{\rm N}}}_{*r}$).

\begin{figure}[ht]
\setlength{\abovecaptionskip}{2mm}
\vspace{-2mm}
\centering
\includegraphics[scale = 0.27]{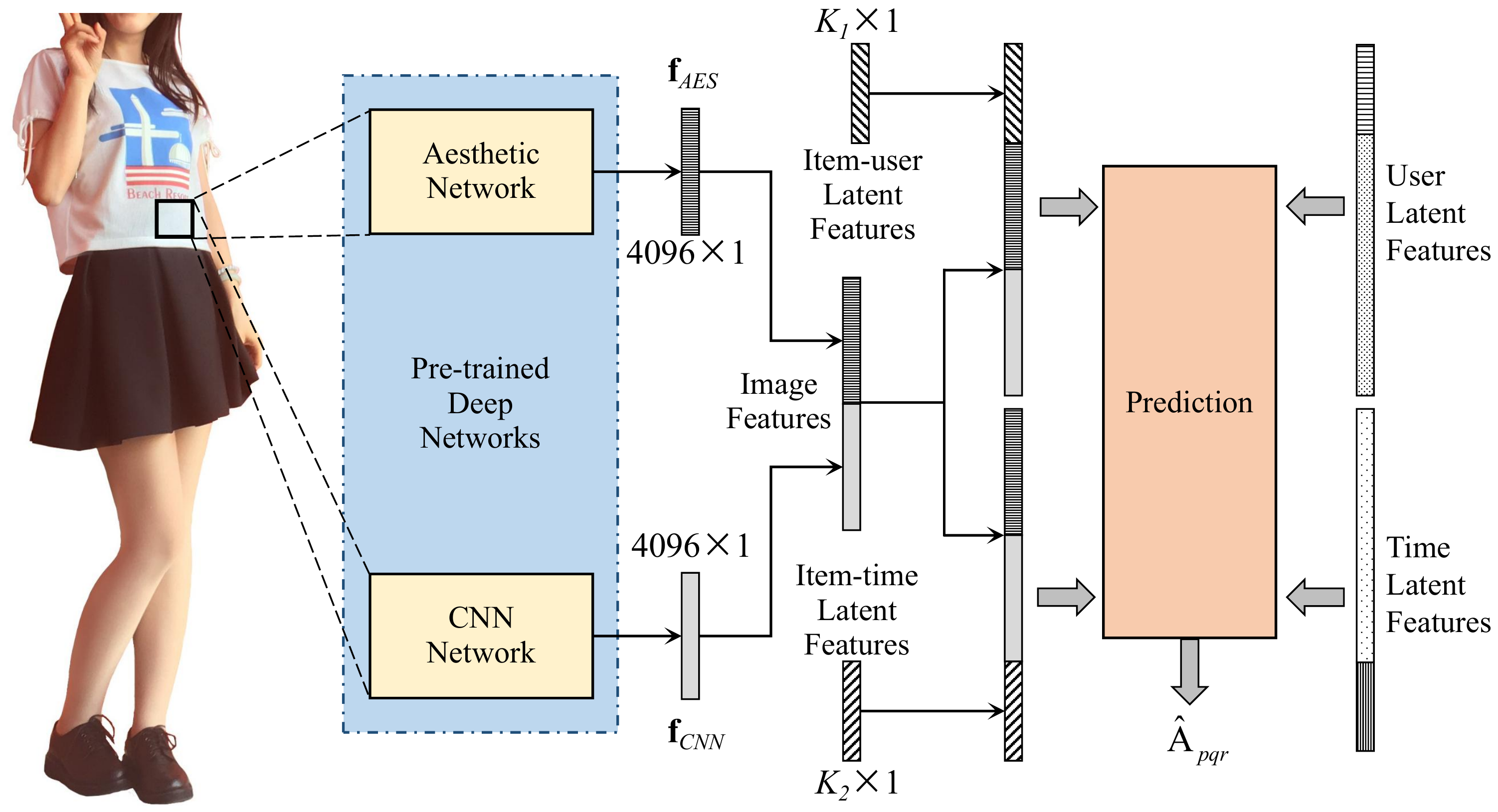}
\caption{Diagram of our preference predictor.}
\label{fig:predictor}
\vspace{-3mm}
\end{figure}

\subsubsection{Model Learning}
The model is optimized with BPR optimization criterion from users' \emph{implicit feedback} (purchase record) with mini-batch gradient descent, which calculates the gradient with a small batch of samples. BPR is a pairwise ranking optimization framework and we represent the training set $D$ into three different forms: $$D_{pr} = \{(p,q,q',r)|p\in \mathcal{P} \,\wedge\, r\in \mathcal{R} \,\wedge\, q\in \mathcal{Q}^+_{pr} \,\wedge\, q'\in \mathcal{Q} \setminus \mathcal{Q}^+_{pr}\},$$ $$D_{p} = \{(p,q,q')|p\in \mathcal{P} \,\wedge\, q\in \mathcal{Q}^+_{p} \,\wedge\, q'\in \mathcal{Q} \setminus \mathcal{Q}^+_{p}\},$$ $$D_{r} = \{(r,q,q')|r\in \mathcal{R} \,\wedge\, q\in \mathcal{Q}^+_{r} \,\wedge\, q'\in \mathcal{Q} \setminus \mathcal{Q}^+_{r}\},$$ where $u$ denotes the user, $r$ represents the time, $q$ represents the positive feedback, and $q'$ represents the non-observed item. The objective function is formulated as:

\begin{flalign}
\label{equ:objective_function2}
\rm BPR\_O\scriptstyle PT = 
&\sum_{(p,q,q',r)\in D_{pr}} \!\! {\rm ln} \, \sigma (\hat{\bm{{\rm A}}}_{pqq'r}) + 
\lambda_1 \!\! \sum_{(p,q,q')\in D_{p}} \!\! {\rm ln} \, \sigma (\hat{\bm{{\rm B}}}_{pqq'})  \nonumber\\
+&\lambda_2 \!\! \sum_{(r,q,q')\in D_{r}} \!\! {\rm ln} \, \sigma (\hat{\bm{{\rm C}}}_{rqq'}) -
{\bm \lambdaup}_{\rm \bm \Theta} {\left\Arrowvert {\rm \bm \Theta} \right\Arrowvert}_{\rm F}^2 \, ,
\end{flalign}
where $\hat{\bm{{\rm A}}}$ is defined in the Equation (\ref{equ:pair_image}), $\hat{\bm{{\rm B}}} = {\bm{{\rm U}}}^\mathsf{T} {\bm{{\rm V}}} + {\bm{{\rm M}}}^\mathsf{T} {\bm{{\rm F}}}$, and $\hat{\bm{{\rm C}}} = {\bm{{\rm T}}}^\mathsf{T} {\bm{{\rm W}}} + {\bm{{\rm N}}}^\mathsf{T} {\bm{{\rm F}}}$; $\hat{\bm{{\rm A}}}_{pqq'r} = \hat{\bm{{\rm A}}}_{pqr} - \hat{\bm{{\rm A}}}_{pq'r}$, $\hat{\bm{{\rm B}}}_{pqq'} = \hat{\bm{{\rm B}}}_{pq} - \hat{\bm{{\rm B}}}_{pq'}$, $\hat{\bm{{\rm C}}}_{rqq'} = \hat{\bm{{\rm C}}}_{rq} - \hat{\bm{{\rm C}}}_{rq'}$; $\sigma$ is the sigmoid function; ${\rm \bm \Theta} = \{{\bm{{\rm U}}}, {\bm{{\rm V}}}, {\bm{{\rm T}}}, {\bm{{\rm W}}}, {\bm{{\rm M}}}, {\bm{{\rm N}}}\}$ and ${\bm \lambdaup}_{\rm \bm \Theta} = \{\lambda_3, \dots, \lambda_8\}$ respectively. We then calculate the gradient of Equation (\ref{equ:objective_function2}). To maximize the objective function, we take the first-order derivatives with respect to each model parameter:

\begin{flalign}
\label{derivatives}
{\nabla_{\bm{{\rm \bm \Theta}}}{\rm BPR\_O\scriptstyle PT}} = &\sigma(-\hat{\bm{{\rm A}}}_{pqq'r})\frac{\partial \hat{\bm{{\rm A}}}_{pqq'r}}{\partial {\rm \bm \Theta}} + \lambda_1 \sigma(-\hat{\bm{{\rm B}}}_{pqq'})\frac{\partial \hat{\bm{{\rm B}}}_{pqq'}}{\partial {\rm \bm \Theta}} \nonumber\\
+ &\lambda_2 \sigma(-\hat{\bm{{\rm C}}}_{rqq'})\frac{\partial \hat{\bm{{\rm C}}}_{rqq'}}{\partial {\rm \bm \Theta}} - {\bm \lambdaup}_{\rm \bm \Theta}{\rm \bm \Theta}. 
\end{flalign}
We use $\rm \bm \thetaup$ to denote certain column of $\rm \bm \Theta$. For our DCFA model, the derivatives are:

\begin{equation}
\label{derivatives1}
\frac{\partial \hat{\bm{{\rm A}}}_{pqq'r}}{\partial {\rm \bm \thetaup}} = \left\{
\begin{array}{lcl}
{\hat{\bm{{\rm C}}}_{rq} {\bm{{\rm V}}}_{*q} - \hat{\bm{{\rm C}}}_{rq'} {\bm{{\rm V}}}_{*q'}} &\text{if} &{\rm \bm \thetaup} = {\bm{{\rm U}}}_{*p} \\
{\hat{\bm{{\rm C}}}_{rq} {\bm{{\rm U}}}_{*p} / -\hat{\bm{{\rm C}}}_{rq'} {\bm{{\rm U}}}_{*p}} &\text{if} &{\rm \bm \thetaup} = {\bm{{\rm V}}}_{*q} / {\bm{{\rm V}}}_{*q'} \\
{\hat{\bm{{\rm C}}}_{rq} {\bm{{\rm F}}}_{*q} - \hat{\bm{{\rm C}}}_{rq'} {\bm{{\rm F}}}_{*q'}} &\text{if} &{\rm \bm \thetaup} = {\bm{{\rm M}}}_{*p} \end{array}  
\right.
\end{equation}

\begin{equation}
\label{derivatives2}
\frac{\partial \hat{\bm{{\rm B}}}_{pqq'}}{\partial {\rm \bm \thetaup}} = \left\{
\begin{array}{lcl}
{{\bm{{\rm V}}}_{*q} - {\bm{{\rm V}}}_{*q'}} &\text{if} &{\rm \bm \thetaup} = {\bm{{\rm U}}}_{*p} \\
{{\bm{{\rm U}}}_{*p} / -{\bm{{\rm U}}}_{*p}} &\text{if} &{\rm \bm \thetaup} = {\bm{{\rm V}}}_{*q} / {\bm{{\rm V}}}_{*q'} \\
{{\bm{{\rm F}}}_{*q} - {\bm{{\rm F}}}_{*q'}} &\text{if} &{\rm \bm \thetaup} = {\bm{{\rm M}}}_{*p}
\end{array}  
\right.
\end{equation}
Equations (\ref{derivatives1}) and (\ref{derivatives2}) give the derivatives for ${\rm \bm \Theta} = \{{\bm{{\rm U}}}, {\bm{{\rm V}}}, {\bm{{\rm M}}}\}$, and we can get the similar form for ${\rm \bm \Theta} = \{{\bm{{\rm T}}}, {\bm{{\rm W}}}, {\bm{{\rm N}}}\}$.

\begin{algorithm}[ht]
\caption{Mini-batch gradient descent based algorithm.}
\LinesNumbered 
\KwIn{sparse tensor $\bm{{\rm A}}$, coupled matrices $\bm{{\rm B}}$ and $\bm{{\rm C}}$, image features $\bm{{\rm F}}$, regularization coefficients ${\rm \bm \lambdaup_{\bm\Theta}}$, batch size $b$, learning rate $\eta$, maximum number of iterations $iter\_max$, and convergence criteria.}
\KwOut{top-$n$ prediction given by the complete tensor $\hat{\bm{{\rm A}}}$.}
initialize $\bm{{\rm \Theta}}$ randomly\;
$iter = 0$\;
\While{not converged $\&\&$ $iter < iter\_max$}{
	$iter += 1$\;
    split all purchase records into $b$-size batches\;
    \For{each batch}{
    	\For{each record in current batch}{
        	select 5 non-observed items $q'$ randomly from $\mathcal{Q} \setminus(\mathcal{Q}^+_{p} \bigcup \mathcal{Q}^+_{r})$\;
            add these negative samples to the current batch\;
        }
        calculate $\nabla_{\bm{{\rm \Theta}}} \rm BPR\_O\scriptstyle PT$ with current batch\;
        $\bm{{\rm \Theta}} = \bm{{\rm \Theta}}+\eta \nabla_{\bm{{\rm \Theta}}} \rm BPR\_O\scriptstyle PT$\;   
    }
    calculate $\hat{{\bm{{\rm A}}}}$ and predict the top-$n$ items\;
}
\Return{the top-$n$ items}\;
\end{algorithm}

We exploit the mini-batch gradient descent to maximize the objective function. For each iteration, all positive samples are enumerated (lines 3-12). We compute the gradients with a batch, including $b$ positive samples (line 5) and $5b$ negative samples (lines 7-9) to construct $5b$ preference pairs, and update the parameters (line 11). To calculate the gradients (line 10), we combine Equations (\ref{derivatives}) with (\ref{derivatives1}) and (\ref{derivatives2}). Of special note is that $\frac{\partial \hat{\bm{{\rm A}}}_{pqq'r}}{\partial {\rm \bm \thetaup}}$ in Equation (\ref{derivatives1}) is certain column of $\frac{\partial \hat{\bm{{\rm A}}}_{pqq'r}}{\partial {\rm \bm \Theta}}$ in Equation (\ref{derivatives}), for example, the $p$-th column when ${\rm \bm \thetaup} = {\bm{{\rm U}}}_{*p}$.

\section{Experiment}

In this section, we conduct experiments on real-world datasets to verify the feasibility of our proposed model. We then analyze the experiment results and demonstrate the precision promotion by comparing it with various baselines. We focus on answering the following three key research questions:

\noindent \textbf{RQ1:} How is the performance of our final framework for the clothing recommendation task?

\noindent \textbf{RQ2:} What are the advantages of the aesthetic features compared with conventional image features?

\noindent \textbf{RQ3:} Is it reasonable to transfer the knowledge gained from \emph{AVA}, which is a dataset of photographic competition works, to the clothing aesthetics assessment task? 

\subsection{Experimental Setup}
\subsubsection{Datasets}
We use the \emph{AVA} dataset to train the aesthetic network and use the \emph{Amazon} dataset to train the recommendation models.
\begin{itemize}
\item{\textbf{Amazon clothing:} The \emph{Amazon} dataset \cite{VBPR} is the consumption records from \emph{Amazon.com}. In this paper, we use the \emph{clothing shoes and jewelry} category filtered with \emph{5-score} (remove users and items with less than 5 purchase records) to train all recommendation models. There are 39,371 users, 23,022 items, and 278,677 records in total (after 2010). The sparsity of the dataset is 99.969\%.}

\item{\textbf{Aesthetic Visual Analysis (AVA):} We train the aesthetic network with the \emph{AVA} dataset \citep{AVA}, which is the collection of images and meta-data derived from \emph{DPChallenge.com}. It contains over 250,000 images with aesthetic ratings from 1 to 10, 66 textual tags describing the semantics of images, and 14 photographic styles (complementary colors, duotones, high dynamic range, image grain, light on white, long exposure, macro, motion blur, negative image, rule of thirds, shallow DOF, silhouettes, soft focus, and vanishing point).}
\end{itemize}

\subsubsection{Baselines}
To demonstrate the effectiveness of our model, we adopt the following methods as baselines for performance comparison:
\begin{itemize}
\item{\textbf{Random (RAND):} This baseline ranks items randomly for all users.}

\item{\textbf{Most Popular (MP):} This baseline ranks items according to their popularity and is non-personalized.}

\item{\textbf{MF:} This \textbf{M}atrix \textbf{F}actorization method ranks items according to the prediction provided by a singular value decomposition structure. It is the basis of many state-of-the-art recommendation approaches.}

\item{\textbf{VBPR:} This is a stat-of-the-art visual-based recommendation method \cite{VBPR}. The image features are pre-generated from the product image using the Caffe deep learning framework.}

\item{\textbf{CMTF:} This is a stat-of-the-art context-aware recommendation method \cite{All_at_once}. The tensor factorization is jointly learned with several coupled matrices.}
\end{itemize}

\subsubsection{Experiment Settings}
In the \emph{Amazon} dataset, we remove the record before 2010 and discretize the time by weeks. There are 237 time intervals, the sparsity of the tensor is 99.99987\%. We randomly split the dataset into training (80\%), validation (10\%), and test (10\%) sets. The validation set was used for tuning hyper-parameters and the final performance comparison was conducted on the test set. We do the prediction and recommend the top-$n$ items to consumers. The Recall and the normalized discounted cumulative gain (NDCG) are calculated to evaluate the performance of the baselines and our model. When $n$ is fixed, the Precision is only determined by true positives whereas the Recall is determined by both true positives and positive samples. To give a more comprehensive evaluation, we exhibit the Recall rather than the Precision and $F_1$-score ($F_1$-score is almost determined by the Precision since the Precision is much smaller than the Recall in our experiments). Our experiments are conducted by predicting Top-5, 10, 20, 50, and 100 favorite clothing.

\subsection{Performance of Our Model (RQ1)}

\begin{figure}[ht]
\setlength{\abovecaptionskip}{2mm}
\vspace{0.1mm}
\centering
\subfigure[]{
\includegraphics[scale = 0.25]{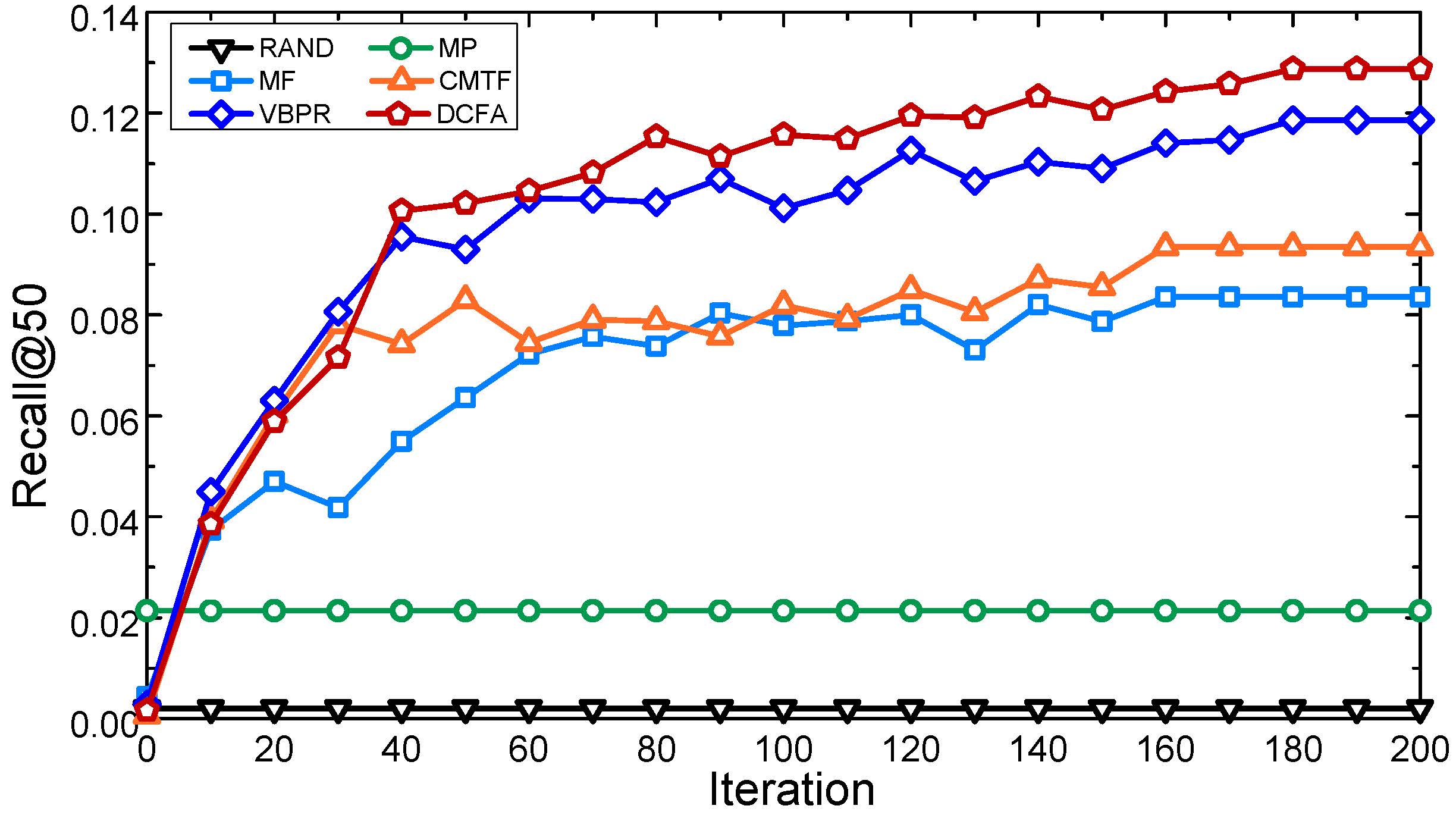}
\label{fig:Recall1}
}
\subfigure[]{
\includegraphics[scale = 0.25]{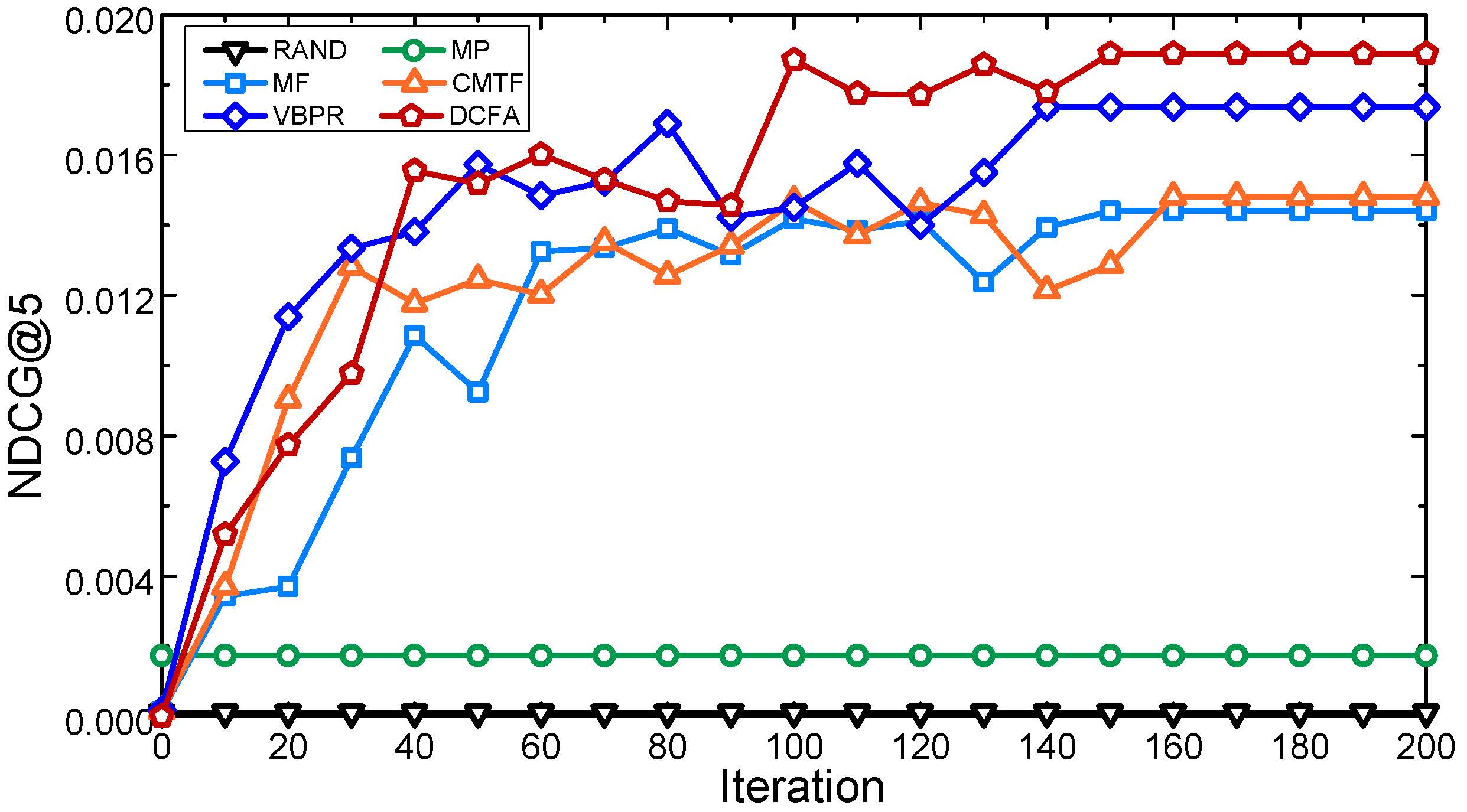}
\label{fig:NDCG1}
}
\caption{Performance with training iterations (test set)}
\label{fig:figure4}
\vspace{-6mm}
\end{figure}

\begin{figure}[ht]
\setlength{\abovecaptionskip}{2mm}
\centering
\subfigure[]{
\includegraphics[scale = 0.25]{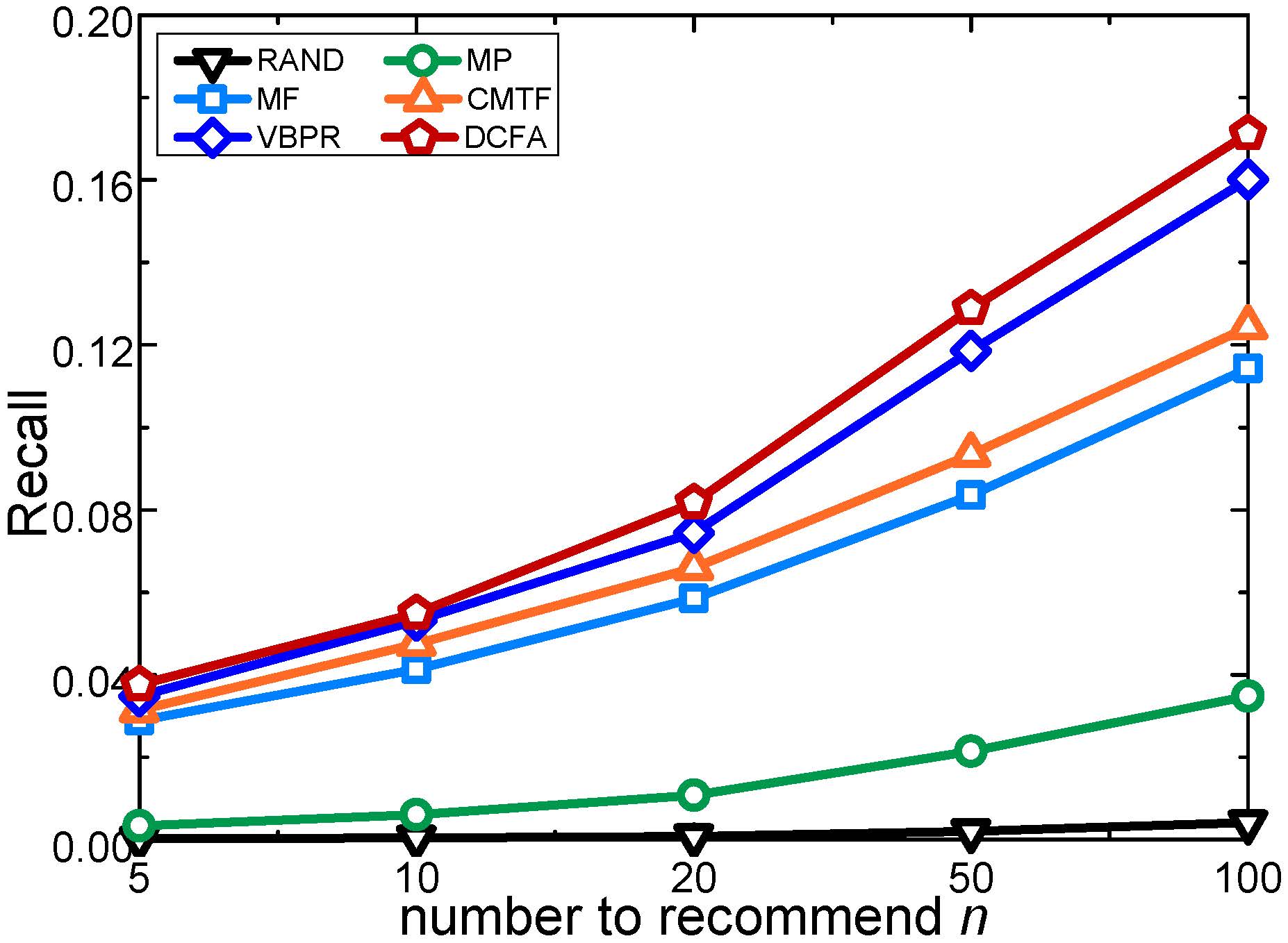}
\label{fig:Recall2}
}
\subfigure[]{
\includegraphics[scale = 0.25]{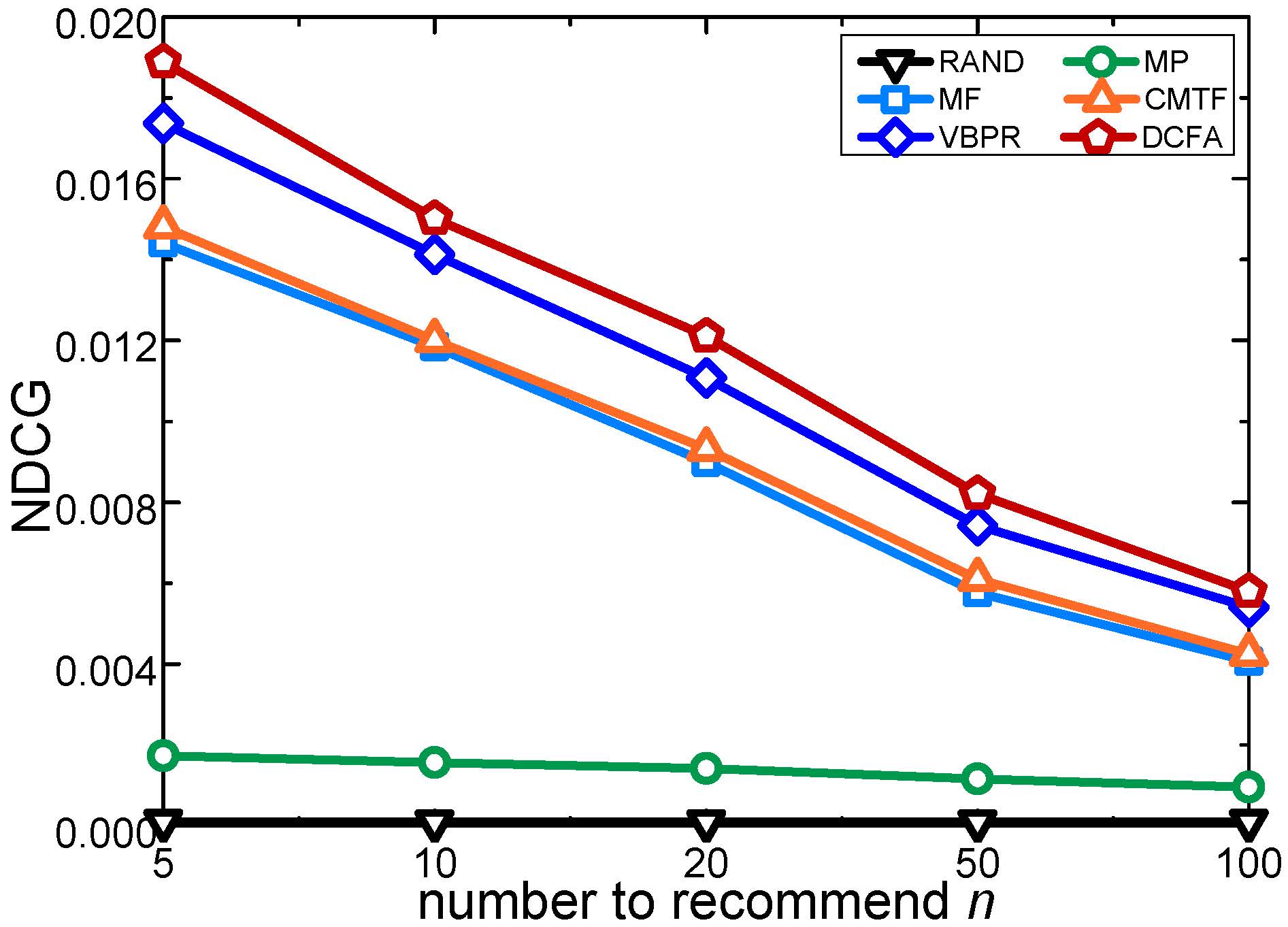}
\label{fig:NDCG2}
}
\caption{Performance with different $n$ (test set)}
\label{fig:figure5}
\vspace{-2mm}
\end{figure}

We iterate 200 times to train all models (except RAND and MP). In each iteration, we enumerate all positive records to optimize models and select 1000 users in test (or validation) set to calculate evaluation metrics, then show the best performance every 10 iterations. Figure \ref{fig:Recall1} shows the Recall and Figure \ref{fig:NDCG1} shows the NDCG during training. We set $n = 50$ when representing the Recall and $n = 5$ when representing the NDCG, due to the relatively large value respectively (represented in Figure \ref{fig:figure5}). We can see that NDCG@5 shows a heavier fluctuation than Recall@50 (Figure \ref{fig:figure4} and Figure \ref{fig:figure6}) since a smaller $n$ leads to a more random prediction. Compared with MP, personalized methods show stronger ability to represent the preference of users and outperform MP several times. By recommending clothes that fit the current season, CMTF can outperform MF on both Recall and NDCG. Enhanced by side information, VBPR performs the best among all baselines. The proposed DCFA model outperforms VBPR about $8.53\%$ on Recall@50 and $8.73\%$ on NDCG@5.

Figure \ref{fig:figure5} represents the variation of the Recall and the NDCG with different $n$. In Figure \ref{fig:Recall2}, we can see that the Recall increases almost linearly with the increasing of $n$ while in Figure \ref{fig:NDCG2}, for most methods (except RAND), the NDCG decreases with the increasing of $n$. Since for most models (except RAND), the higher-rated clothing is with more possibility to be chosen by consumers. So the ordering quality decreases with the increasing of $n$. To the contrary, since RAND orders all items randomly, its ordering quality keeps constant.


\begin{figure*}[ht]
\setlength{\abovecaptionskip}{2mm}
\centering
\subfigure[]{
\includegraphics[scale = 0.175]{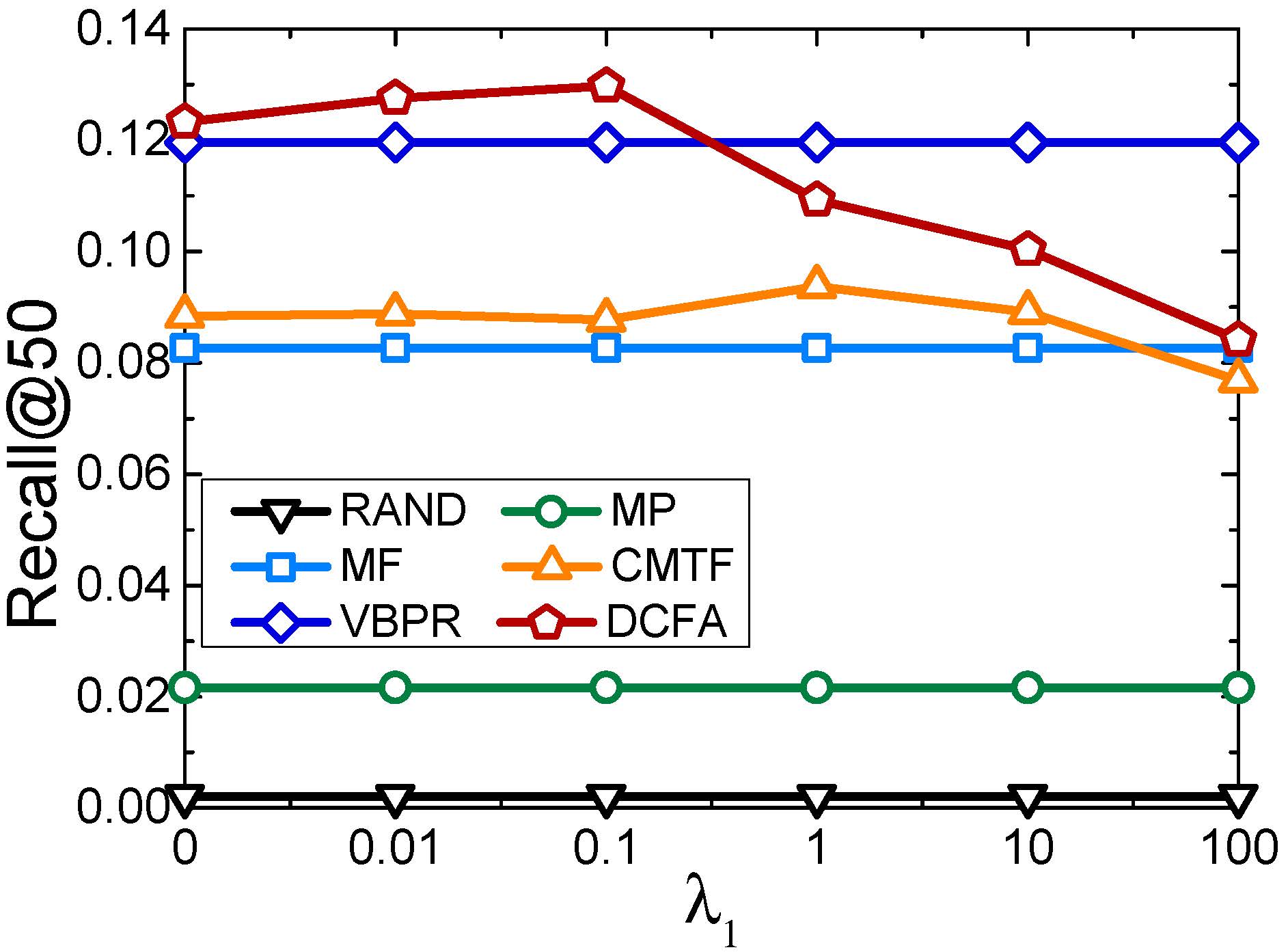}
\label{fig:l1}
}
\subfigure[]{
\includegraphics[scale = 0.175]{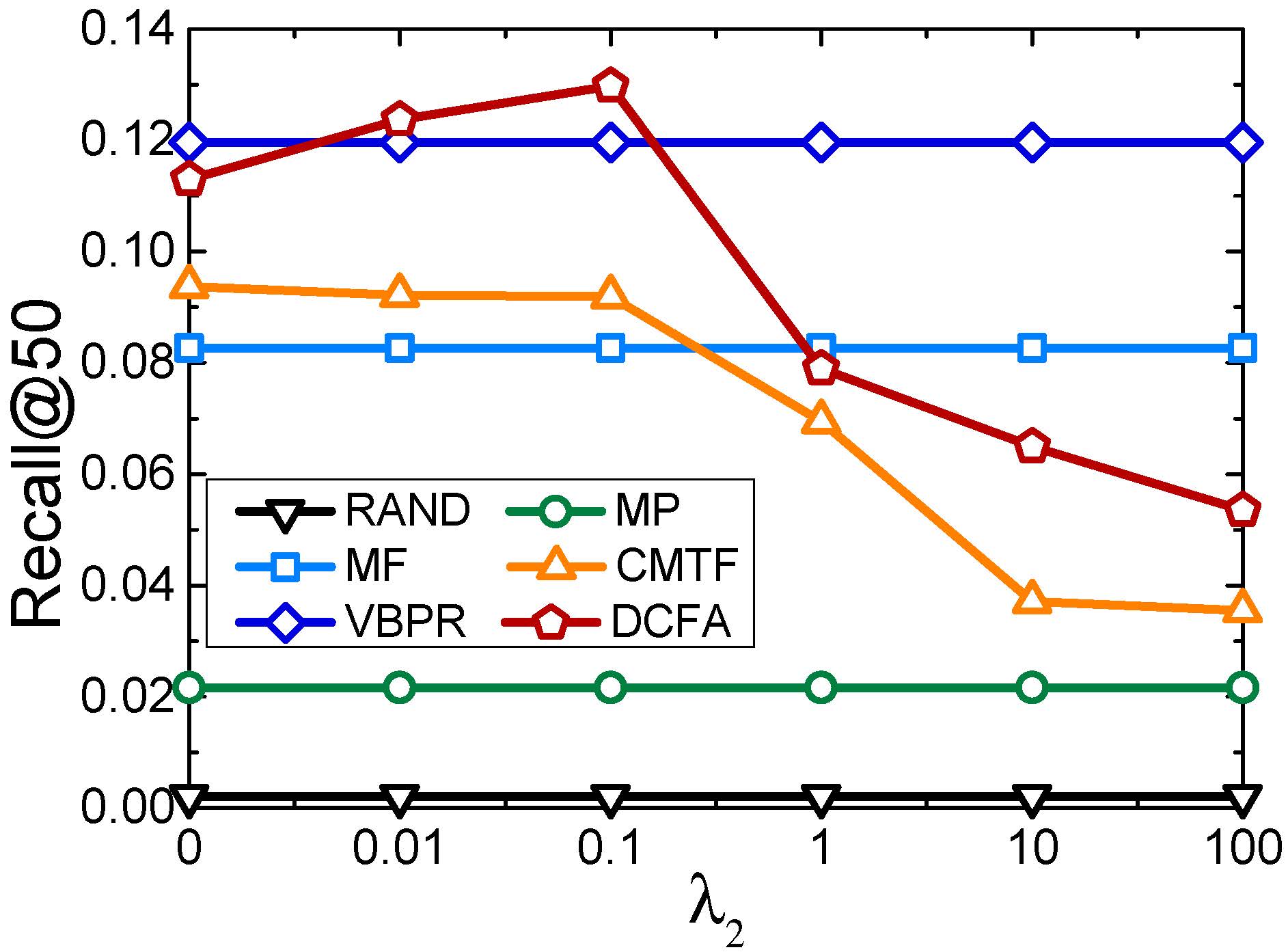}
\label{fig:l2}
}
\subfigure[]{
\includegraphics[scale = 0.175]{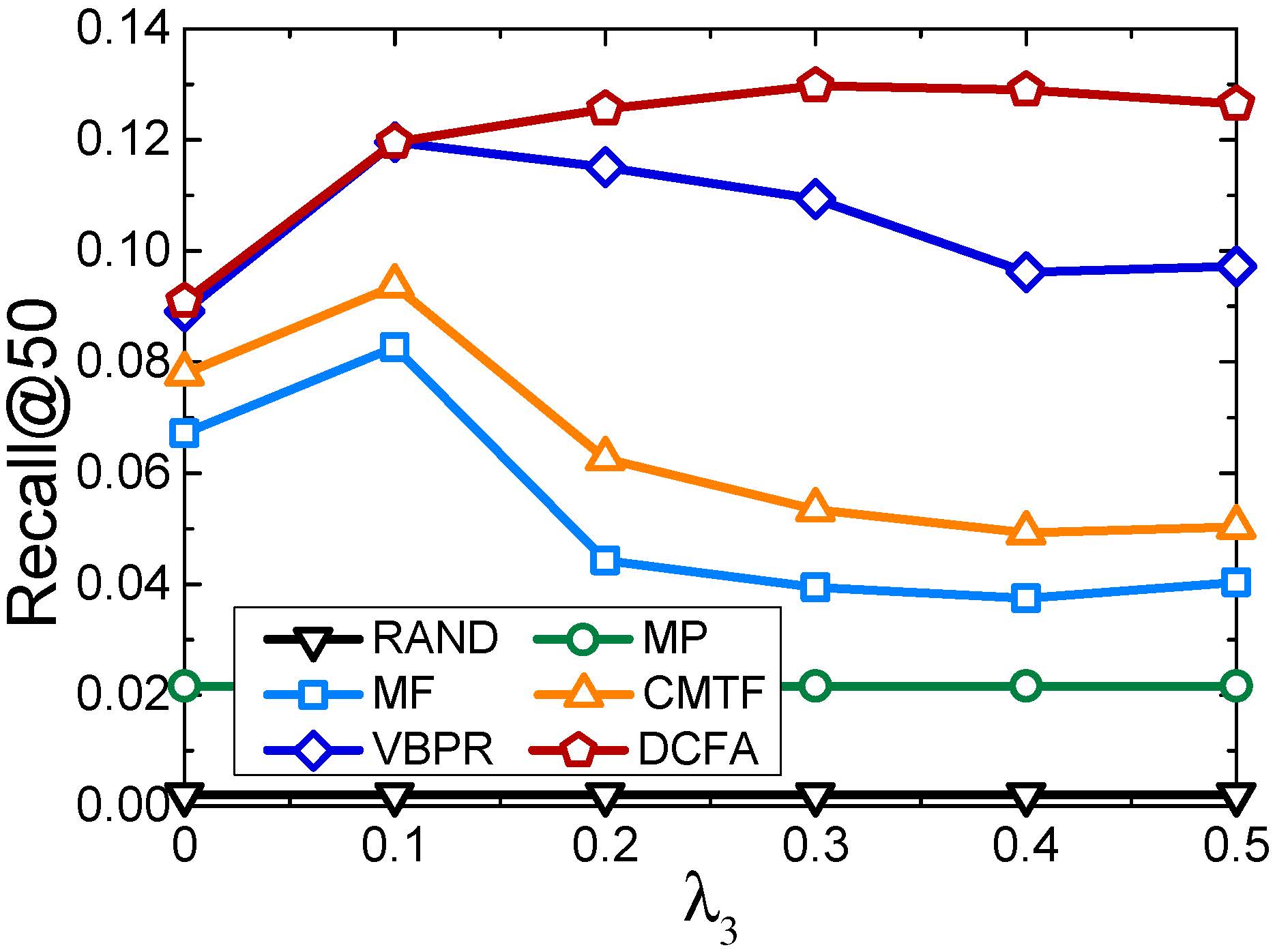}
\label{fig:l3}
}
\subfigure[]{
\includegraphics[scale = 0.175]{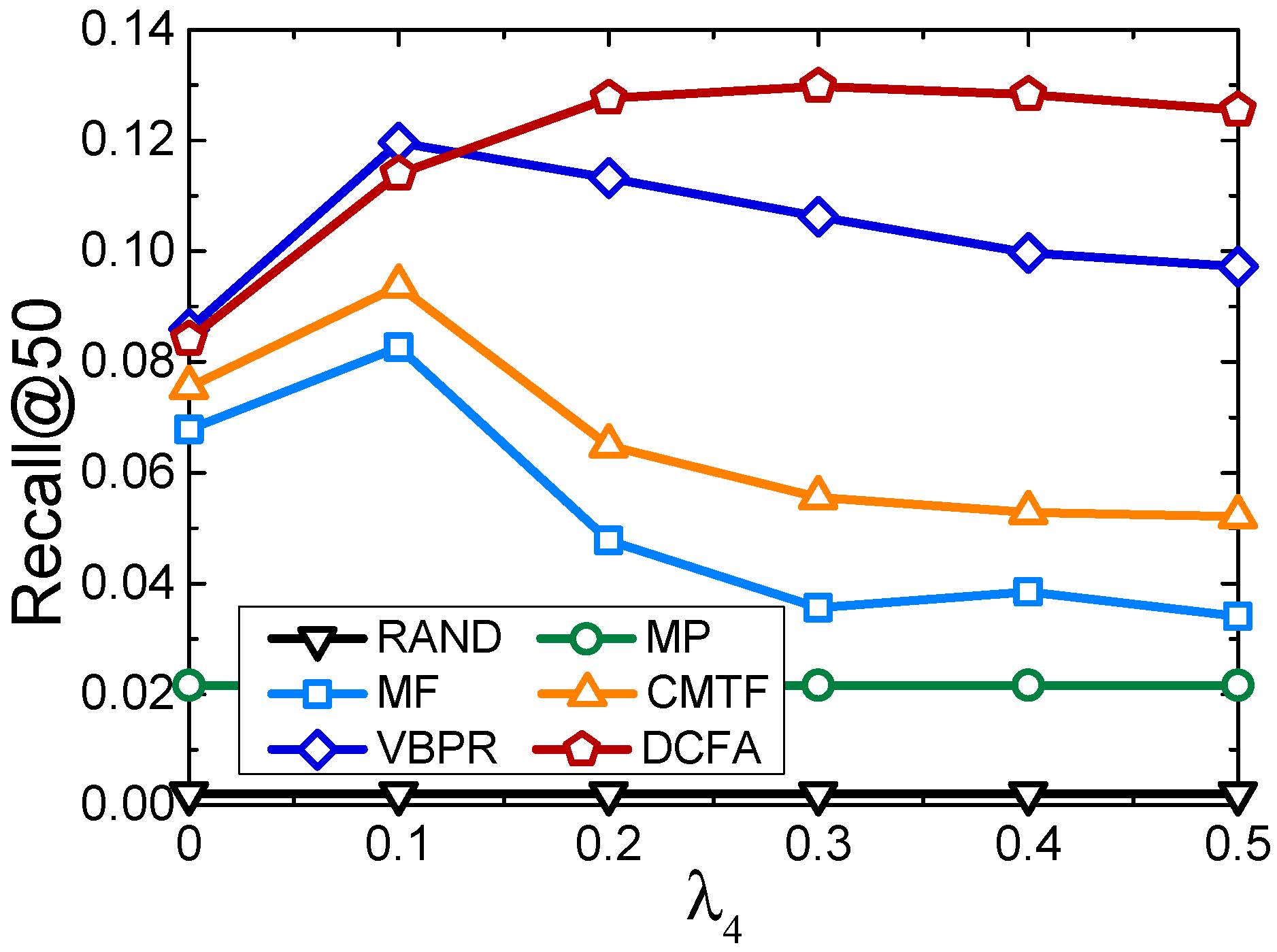}
\label{fig:l4}
}
\subfigure[]{
\includegraphics[scale = 0.175]{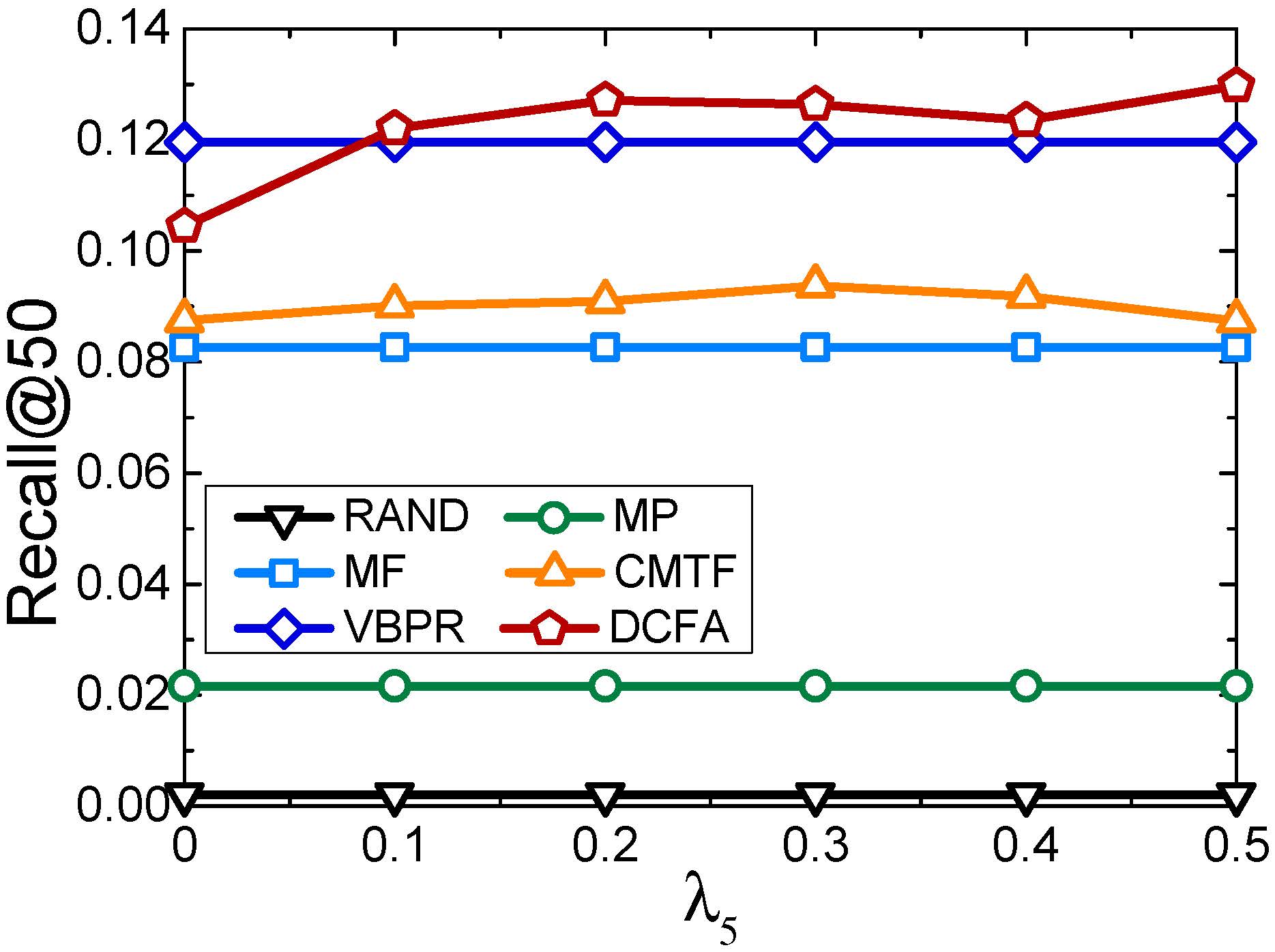}
\label{fig:l5}
}
\subfigure[]{
\includegraphics[scale = 0.175]{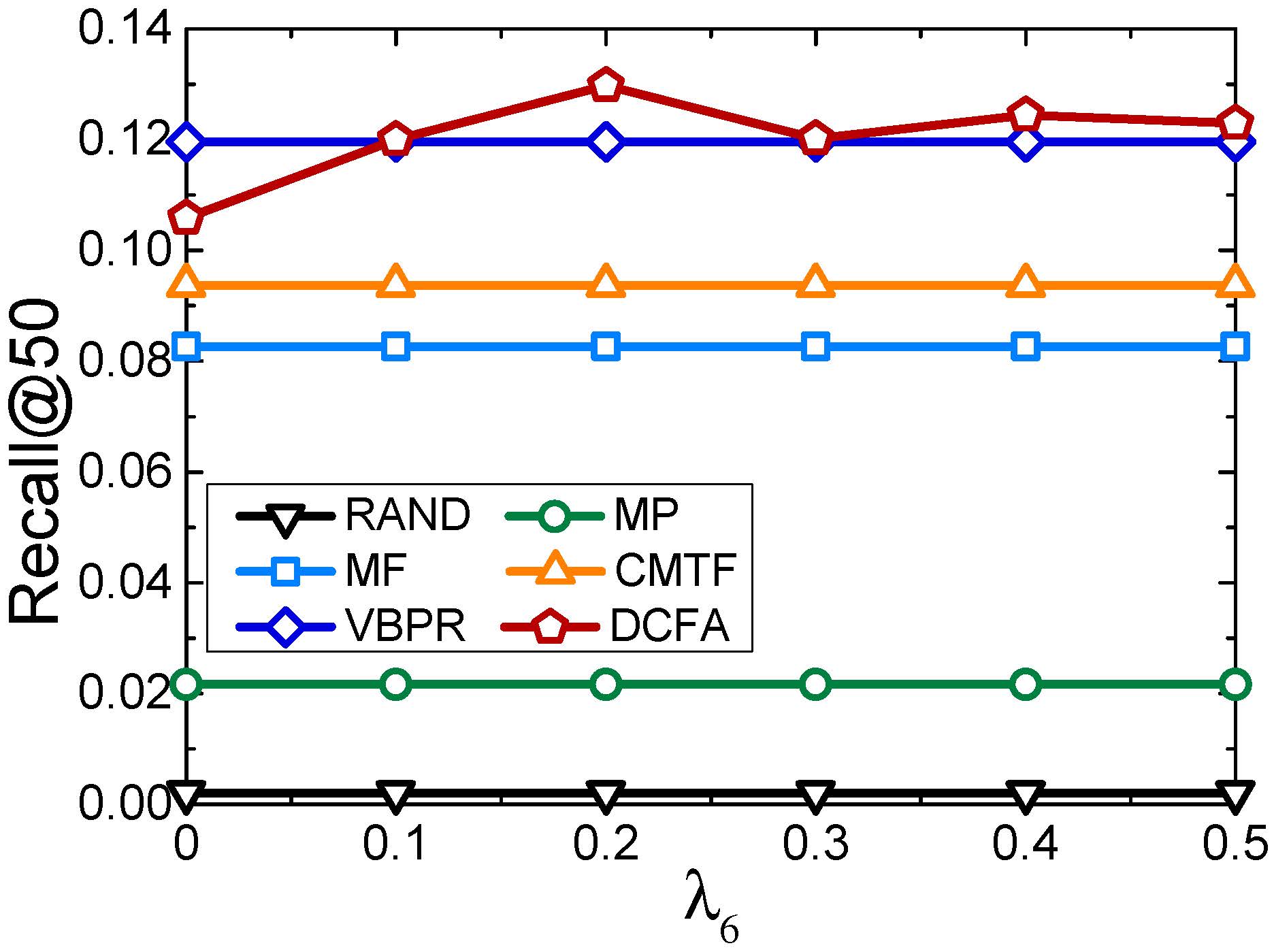}
\label{fig:l6}
}
\subfigure[]{
\includegraphics[scale = 0.175]{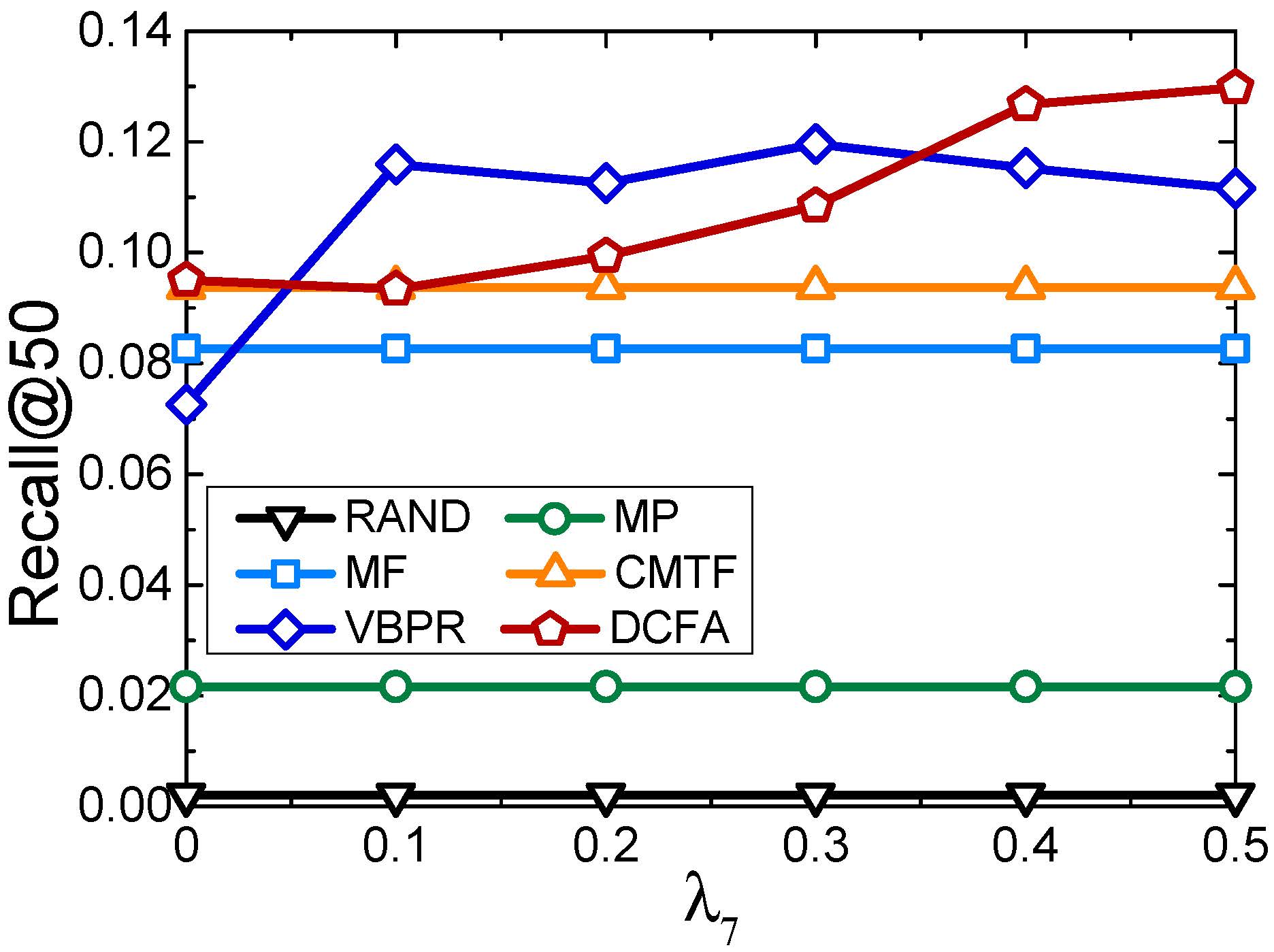}
\label{fig:l7}
}
\subfigure[]{
\includegraphics[scale = 0.175]{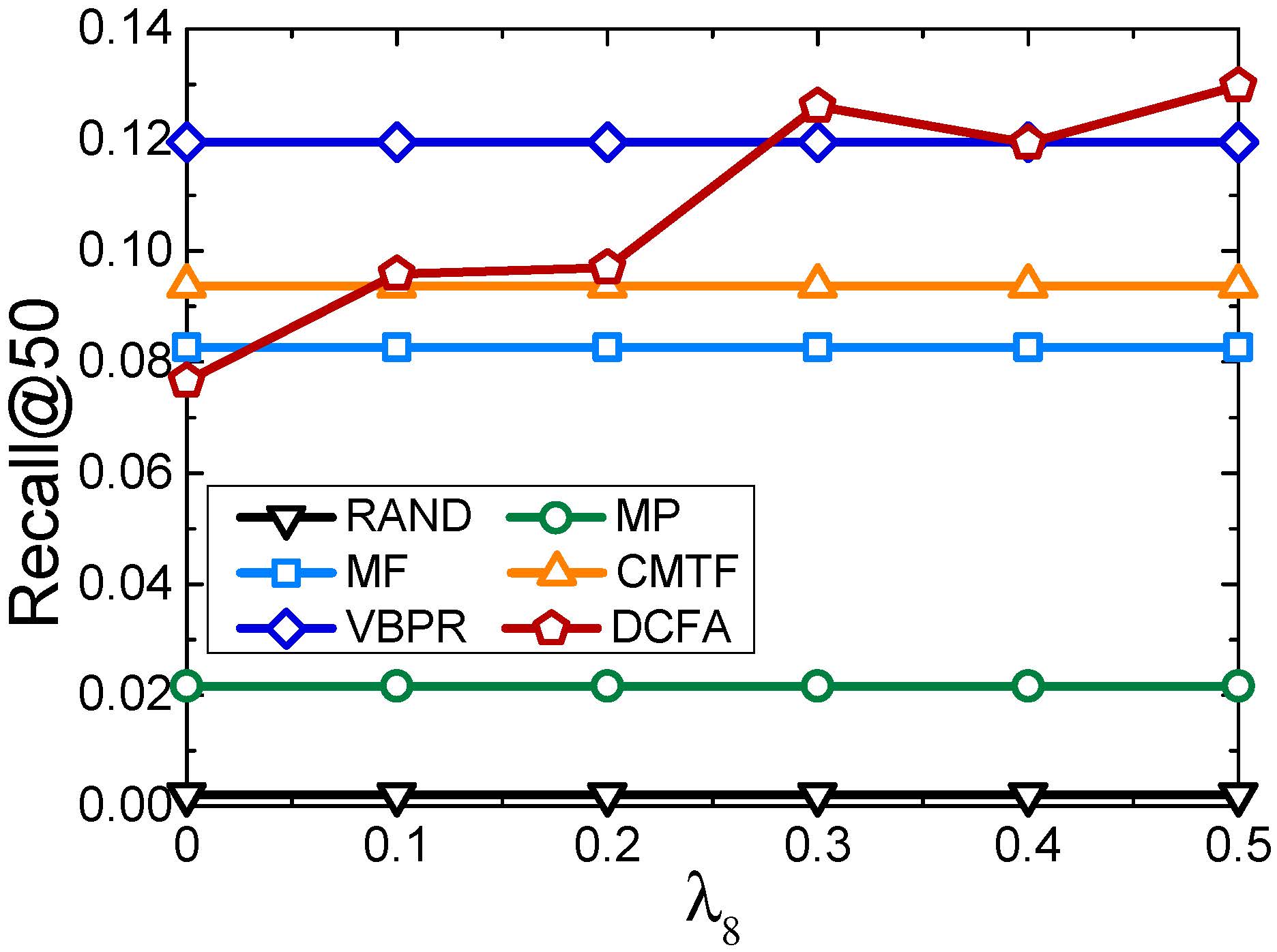}
\label{fig:l8}
}

\caption{Impacts of hyperparameters (validation set)}
\label{fig:f6}
\end{figure*}

In our experiments, we tune all hyperparameters sequentially on the validation set (include those in our model and in baselines). There are 8 hyperparameters in Equation (\ref{equ:objective_function2}) and the sensitivity analysis is shown in Figure \ref{fig:f6}. We can see that when $\lambda_1 = 0.1$, $\lambda_2 = 0.1$, $\lambda_3 = 0.3$, $\lambda_4 = 0.3$, $\lambda_5 = 0.5$, $\lambda_6 = 0.2$, $\lambda_7 = 0.5$, $\lambda_8 = 0.5$, DCFA can achieve the best performance. Influences of hyperparameters in baselines are also shown in Figure \ref{fig:f6}. For all models, $\lambda_1$ and $\lambda_2$ are used to represent weights of the coupled user-item matrix and time-item matrix. $\lambda_3$ to $\lambda_8$ are regularization coefficients of the user matrix, item matrix (connecting with user), time matrix, item matrix (connecting with time), aesthetic preference matrix of consumers, and aesthetic preference matrix of time respectively. For example, we can see that the performance of MF varies with regularization coefficients of the user matrix ($\lambda_3$) and the item matrix (connecting with user, $\lambda_4$), while keeps constant with the variation of $\lambda_5$ because there is no time matrix in MF. Specially, in CMTF, the item matrix connects both the user and time matrices, we use $\lambda_3$, $\lambda_4$, $\lambda_5$ to represent the regularization coefficients of the user, item, and time matrices respectively.

\subsection{Necessity of the aesthetic features (RQ2)}

In this subsection, we discuss the necessity of the aesthetic features. We combine various widely used features to our basic model and compare the effect of each features by constructing five models:

\begin{itemize}
\item{\textbf{DCF:} This is our basic \textbf{D}ynamic \textbf{C}ollaborative \textbf{F}iltering model without any image features, which is represented in the subsection 4.1.}

\item{\textbf{DCFH:} This is a \textbf{D}ynamic \textbf{C}ollaborative \textbf{F}iltering model with Color \textbf{H}istograms.}

\item{\textbf{DCFCo:} This is a \textbf{D}ynamic \textbf{C}ollaborative \textbf{F}iltering model with \textbf{C}NN Features \textbf{o}nly.}

\item{\textbf{DCFAo:} This is a \textbf{D}ynamic \textbf{C}ollaborative \textbf{F}iltering model with \textbf{A}esthetics Features \textbf{o}nly.}

\item{\textbf{DCFA:} This is our proposed model represented in the subsection 4.2, utilizing both CNN features and aesthetic features.}
\end{itemize}

\begin{figure}[ht]
\setlength{\abovecaptionskip}{2mm}
\centering
\subfigure[]{
\includegraphics[scale = 0.171]{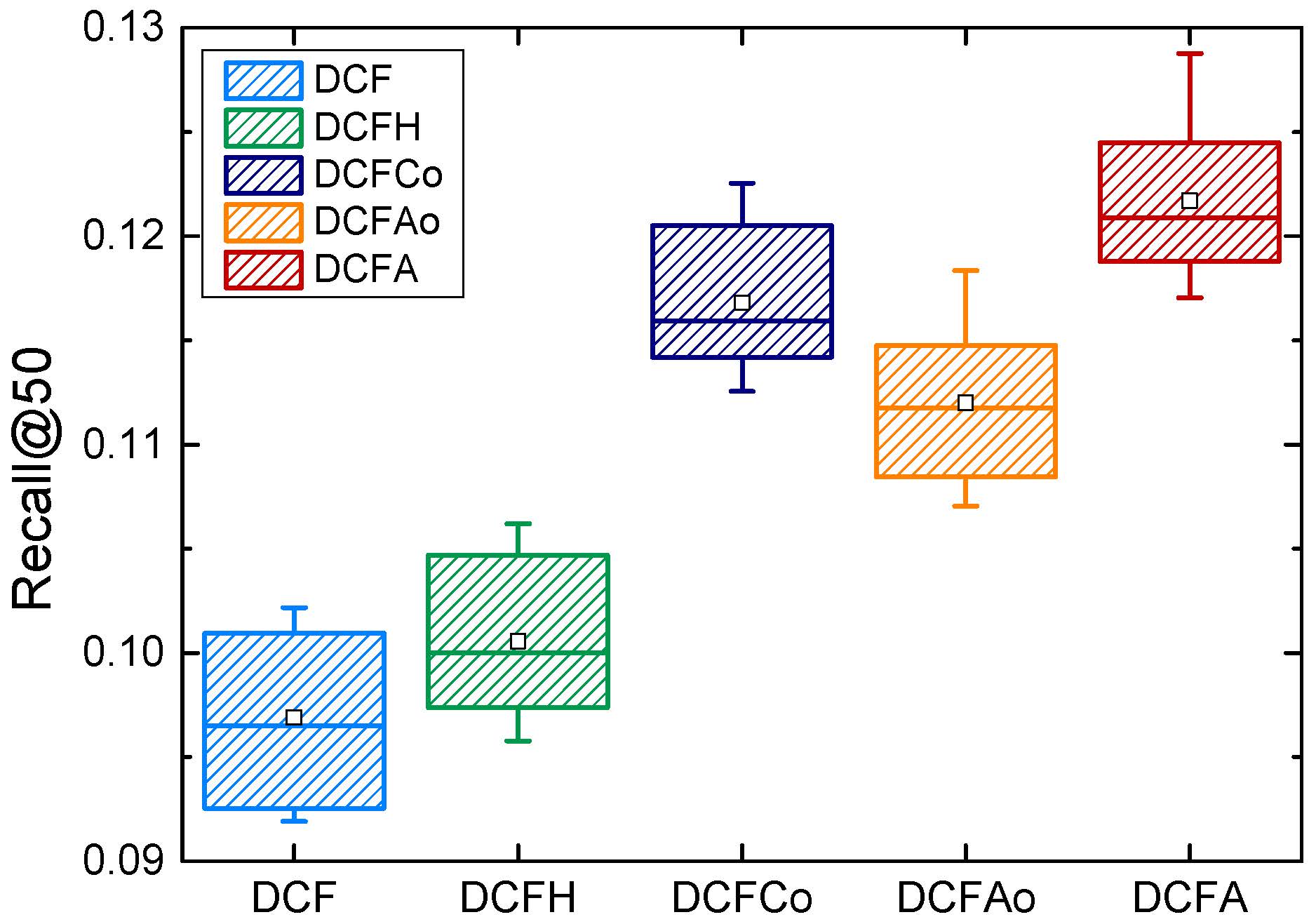}
\label{fig:Recall3}
}
\hspace{-2mm}
\subfigure[]{
\includegraphics[scale = 0.171]{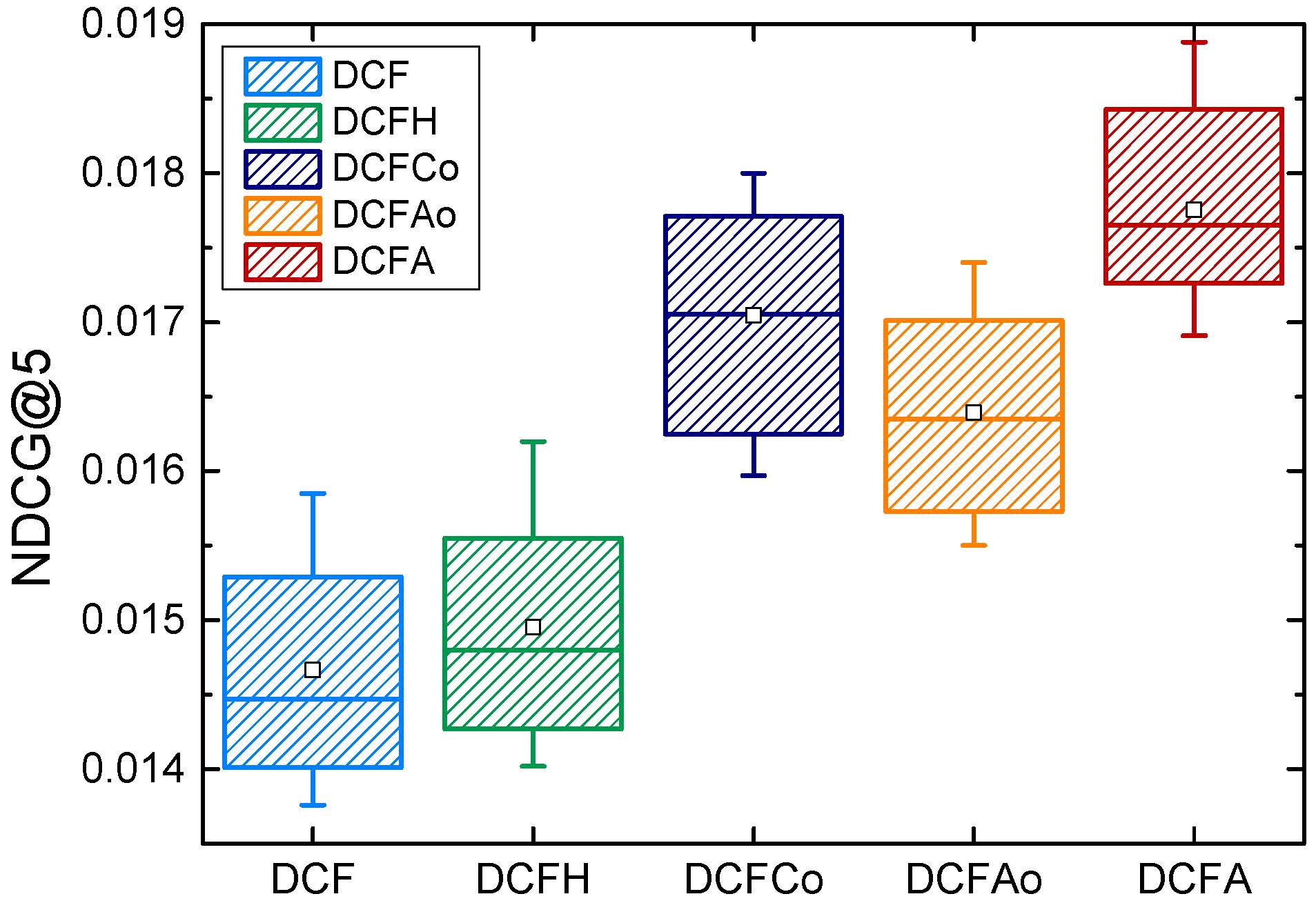}
\label{fig:NDCG3}
}
\caption{Performance of various features (test set)}
\label{fig:figure6}
\vspace{-4mm}
\end{figure}

Figures \ref{fig:Recall3} and \ref{fig:NDCG3} show the distribution of 10 maximum on Recall@50 and the NDCG@5 of each model during the 200 iterations. As shown in Figure \ref{fig:figure6}, DCF performs the worst since no image features are involved to provide the extra information. With the information of color distribution, DCFH performs better, though still worse than DCFCo and DCFAo, because the low-level features are too crude and unilateral, and can provide very limited information about consumers' aesthetic preference. 
DCFCo and DCFAo show the similar performance because both CNN features and aesthetic features have strong ability to mine the user's preference. 
Our DCFA model, capturing both semantic information and aesthetic information, performs the best on the \emph{Amazon} dataset since those two kinds of information mutually enhance each other to a certain extent. Give an intuitive example, if a consumer want to purchase a skirt, she needs to tell whether there is a skirt in the image (semantic information) when look through products, and then she needs to evaluate if the skirt is good-looking and fits her tastes (aesthetic information) to make the final decision. We can see that in the actual scene, semantic information and aesthetic information are both important for decision making and the two kinds of features complement each other in modeling this procedure. Though CNN features also contain some aesthetic information (like color, texture, etc.), it is far from a comprehensive description, which can be provide by the aesthetic features on account of the abundant raw aesthetic features inputted and training for aesthetic assessment tasks. Also, aesthetic features contain some holistic information (like structure and portion), while cannot provide a complete semantic description. So, these two kind of features cannot replace each other and are supposed to model users' preference collaboratively. In our experiments, DCFA outperforms DCFCo and DCFAo about $5.06\%$ and $8.79\%$ on Recall@50, $4.89\%$ and $8.51\%$ on NDCG@5 respectively. We can see that though the aesthetic features and CNN features do not perform the best separately, they mutually enhance each other and achieve improvement together.

\begin{figure*}[ht]
  \centering
  \subfigure[]{
    \label{fig:fa} 
    \includegraphics[scale = 0.58]{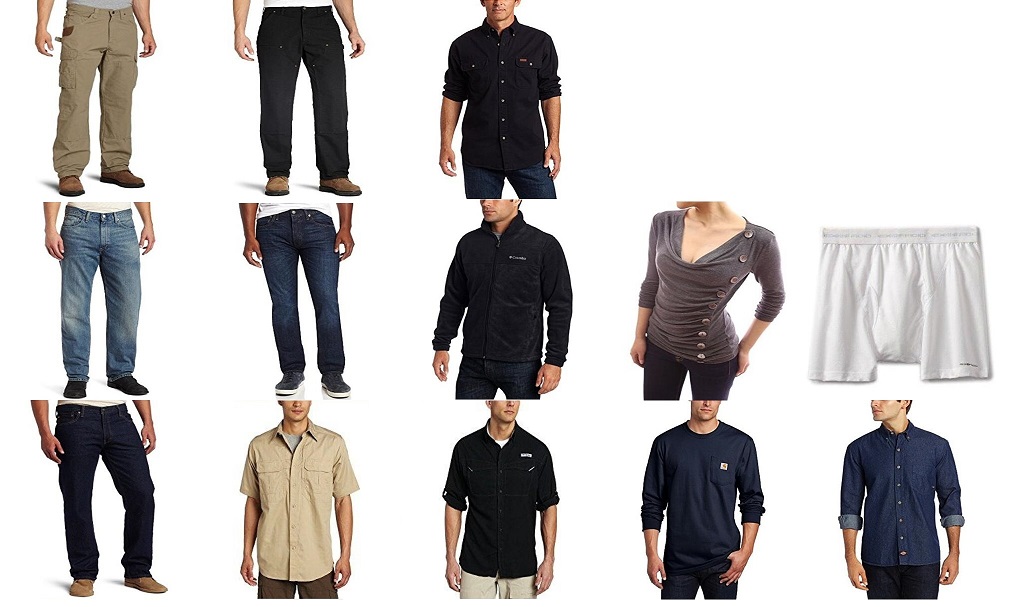}}
  \subfigure[]{
    \label{fig:fb} 
    \includegraphics[scale = 0.58]{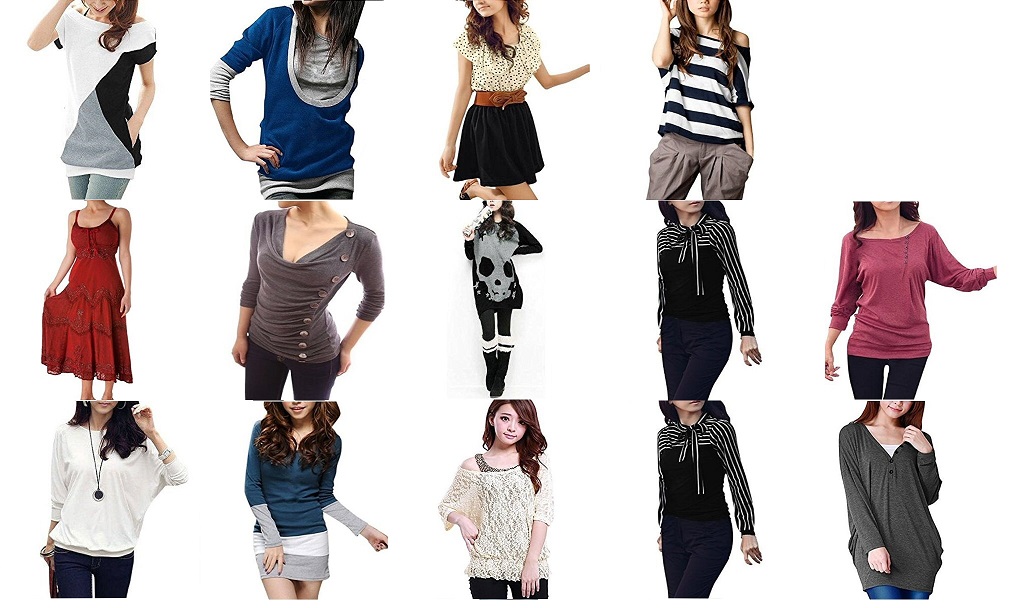}}
  \subfigure[]{
    \label{fig:fc} 
    \includegraphics[scale = 0.58]{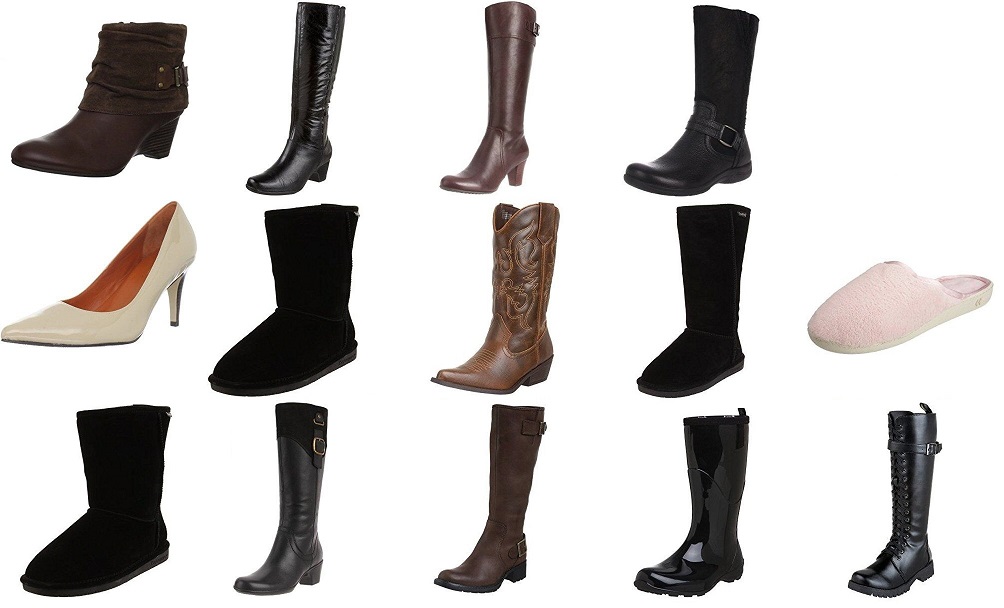}}
  \subfigure[]{
    \label{fig:fd} 
    \includegraphics[scale = 0.58]{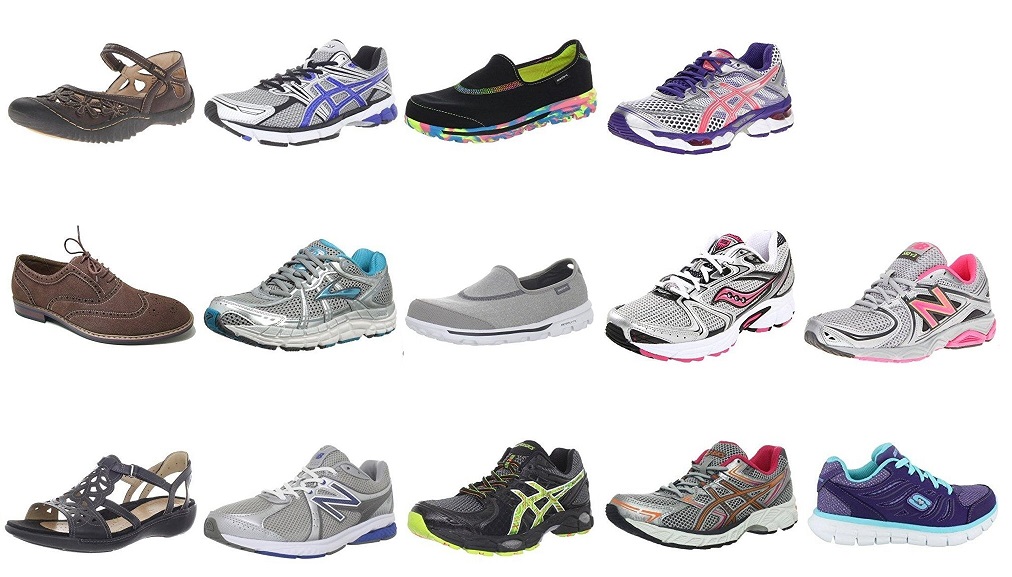}}
  \subfigure[]{
    \label{fig:fe} 
    \includegraphics[scale = 0.58]{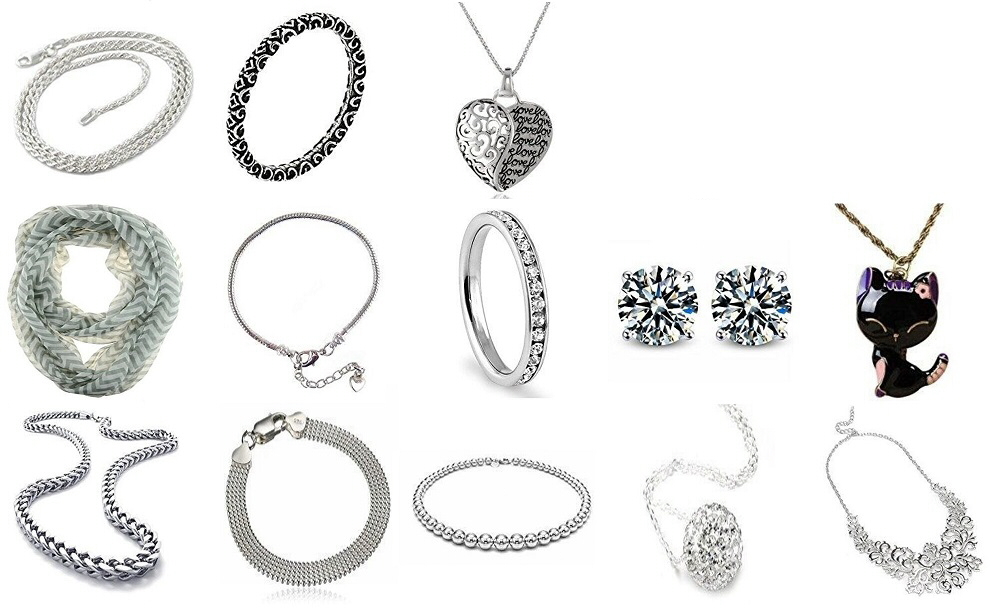}}
  \subfigure[]{
    \label{fig:ff} 
    \includegraphics[scale = 0.58]{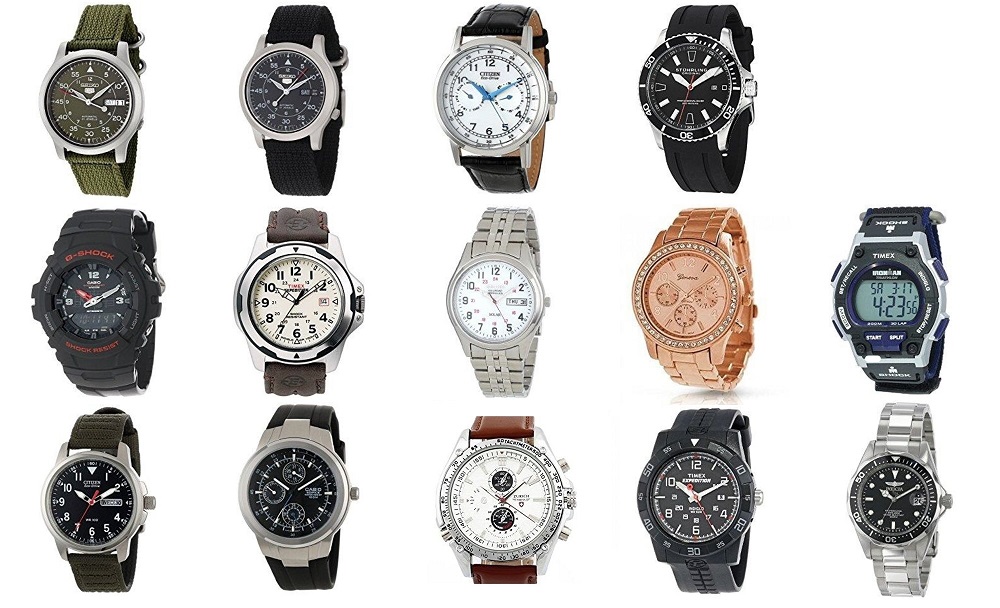}}
  \caption{Items purchased by consumers and recommended by different models.}
  \label{fig:compare} 
\end{figure*}

Several purchased and recommended items are represented in the Figure \ref{fig:compare}. Items in the first row are purchased by certain consumer (training data, the number is random). To illustrate the effect of the aesthetic features intuitively, we choose the consumers with explicit style preference and single category of items. Items in the second row and third row are recommended by DCFCo and DCFA respectively. For these two rows, we choose five best items from the 50 recommendations to exhibit. Comparing the first and the second row, we can see that leveraging semantic information, DCFCo can recommend the congeneric (with the CNN features) and relevant (with tensor factorization) commodities. Though can it recommend the pertinent products, they are usually not in the same style with what the consumer has purchased. Capturing both aesthetic and semantic information, DCFA performs much better. We can see that items in the third row have more similar style with the training samples than items in the second row. Take Figure \ref{fig:ff} as an example, we can see that what the consumer likes are vibrant watches for young men. However, watches in the second row are in pretty different styles, like digital watches for children, luxuriantly-decorated ones for ladies, old-fashioned ones for adults. Evidently, watches in the third row are in similar style with the train samples. They have similar color schemes and design elements, like the intricatel-designed dials, nonmetallic watchbands, small dials, and tachymeters. It is also obvious in Figure \ref{fig:fc}, we can see that the consumer prefers boots, ankle boots or thigh boots. However, products recommended by DCFCo are some different type of women's shoes, like high heels, snow boots, thigh boots, and cotton slippers. Though there is a thigh boot, it is not in line with the consumer's aesthetics due to the gaudy patterns and stumpy proportion, which rarely appears in her choices. Products recommended by DCFA are better. First, almost all recommendations are boots. Then, thigh boots in the third row are in the same style with the training samples, like leather texture, slender proportions, simple design and some design elements of detail like straps and buckles (the second and third ones). Though the last one seems a bit different with the training samples, it is in the uniform style with them intuitively, since they are all designed for young ladies. As we can see, with the aesthetic features and the CNN features complementing each other, DCFA performs much better than DCFCo.

\vspace{-3mm}
\subsection{Rationality of using the \emph{AVA} dataset (RQ3)}

The BDN is trained on the \emph{AVA} dataset, which contains photographic works labeled with aesthetic ratings, textual tags, and photographic styles. We utilize aesthetic ratings and photographic styles to train the aesthetic network. In this subsection, we simply discuss if it is reasonable to estimate clothing by the features trained for photographic assessment.

With no doubt that there are many similarities between esthetical photographs and well-designed clothing, like delightful color combinations, saturation, brightness, structures, proportion, etc. Of course, there are also many differences. To address this gap, we modify the BDN. In \cite{Brain}, there are 14 pathways to captures all photographic styles. In this paper, we remove several pathways for the photographic styles which contribute little in clothing estimation, like high dynamic range, long exposure, macro, motion blur, shallow DOF, and soft focus. These features mainly describe the camera parameters setting or photography skills but not the image, so they help little in our clothing aesthetic assessment task. Experiments show that our proposed model can uncover consumers' aesthetic preference and recommend the clothing that are in line with their aesthetics, and the performance is obviously promoted.

There are many works recommending clothing or garments with fashion information \cite{clothes1,clothes2,clothes3} and there are several datasets for clothing fashion style. \cite{clothes2} utilized three datasets containing street fashion images and annotations by fashionistas to train phase, input queries, and return ranked list respectively. \cite{clothes3} proposed a novel dataset crawled from \emph{chictopia.com} containing photographs, text in the form of descriptions, votes, and garment tags. However, these datasets are mainly for fashion style and not appropriate for BDN training because of the lack of aesthetic ratings and style tags, so we choose \emph{AVA}. There are abundant images and tags to provide raw aesthetic features. Though not all raw features are needed due to the gap of photographic works and clothing, many of them
are important in clothing aesthetic assessment. Beyond that, our model should have ability to extend to a wider range of application scenarios, like the recommendation of electronic products, movies, toys, etc., so a general dataset for aesthetic network training is important.

\vspace{-2mm}
\section{Conclusion}

In this paper, we investigated the usefulness of aesthetic features for personalized recommendation on implicit feedback datasets. We proposed a novel model that incorporates aesthetic features into a tensor factorization model to capture the aesthetic preference of consumers at a particular time. Experiments on challenging real-word datasets show that our proposed method dramatically outperforms state-of-the-art models, and succeeds in recommending items that fit consumers' style. 

For future work, we will establish a large dataset for product aesthetic assessment, and train the networks to extract the aesthetic information better. Moreover, we will investigate the effectiveness of our proposed method in the setting of explicit feedback. Lastly, we are interested in integrating the domain knowledge about aesthetic assessment, e.g., in the form of decision rules~\cite{he_TEM}, into the recommender model. 
\newpage\balance

\bibliographystyle{ACM-Reference-Format}\balance
\bibliography{ref}\balance

\end{document}